\RequirePackage{fix-cm}
\documentclass[smallextended]{svjour3}       
\smartqed  
\usepackage{graphicx}
\graphicspath{{figures/}}
\DeclareGraphicsExtensions{.eps,.pdf,.png}

\usepackage{natbib}
\setcitestyle{authoryear}

\usepackage[usenames,dvipsnames,svgnames,table]{xcolor} 

\usepackage[export]{adjustbox} 
\usepackage{amssymb,amsmath}

\usepackage[hyphens]{url}

\usepackage{enumitem}

\usepackage{calc}

\usepackage[export]{adjustbox}

\usepackage[framemethod=tikz]{mdframed}
\newmdenv[
  innerleftmargin=7pt,
  innerrightmargin=7pt,
  skipabove=\baselineskip,
  skipbelow=\baselineskip,
  tikzsetting={draw=black,dashed,line width=0.5pt,dash pattern = on 4pt off 2pt},
  linecolor=white,
  backgroundcolor=white
]{dashedbox}
\newmdenv[
  innerleftmargin=7pt,
  innerrightmargin=7pt,
  skipabove=\baselineskip,
  skipbelow=\baselineskip,
  tikzsetting={draw=black, line width=0.5pt},
  linecolor=black,
  backgroundcolor=white
]{normalbox}
\newmdenv[
  topline=false,
  bottomline=false,
  rightline=false,
  skipabove=\topsep,
  skipbelow=\topsep,
  innertopmargin=0pt,
  innerbottommargin=0pt,
  innerleftmargin=7pt,
  innerrightmargin=0pt,
  tikzsetting={draw=black, line width=3pt},
  linecolor=black,
  backgroundcolor=white
]{verticalline}

\usepackage{listings}

\definecolor{light-gray}{gray}{0.97}
\definecolor{gray}{rgb}{0.4,0.4,0.4}
\definecolor{darkblue}{rgb}{0.0,0.0,0.6}
\definecolor{cyan}{rgb}{0.0,0.6,0.6}

\lstnewenvironment{code} {\lstset{
}}{}

\lstnewenvironment{java} {\lstset{
  language=Java
}}{}

\lstnewenvironment{regex} {\lstset{
  alsoletter=\\()\{\}.,
  morekeywords={String, long, if, return, format,B,int,Math,log}
}}{}

\lstdefinelanguage{XML}
{
  morestring=[b]",
  morestring=[s]{>}{<},
  morecomment=[s]{<?}{?>},
  stringstyle=\color{black},
  identifierstyle=\color{darkblue},
  keywordstyle=\color{cyan},
  morekeywords={xmlns,version,type,xmlns:data,xmlns:xlink,height,width,preserveAspectRatio,viewBox,style,data,id,xlink,href,authors,date,time,annotation,artifact,path,x,y,role,rel,inline,iconpos,transition,icon,theme,class,src}
}

\lstnewenvironment{sql} {\lstset{
	morekeywords={SELECT, FROM, WHERE, ORDER, BY, ASC, LIKE, INNER, JOIN, ON} 
}}{}

\usepackage{tabularx}

\usepackage{multirow}

\usepackage{hyperref}

\definecolor{VeryLightGray}{rgb}{0.92,0.92,0.92}
\usepackage{colortbl}
\newcommand{\graybg}{\cellcolor{VeryLightGray}}

\usepackage{enumitem}

\usepackage{calc}

\usepackage{bm} 

\usepackage{nicefrac}

\usepackage{amsmath}
\usepackage{amssymb} 
\usepackage{eucal} 

\usepackage{siunitx}
\usepackage{algpseudocode}
\usepackage{algorithm}

\begin{document}

\title{The Evolution of Stack~Overflow Posts}
\subtitle{Reconstruction and Analysis}


\author{Sebastian Baltes \and Lorik Dumani \and Christoph Treude \and Stephan Diehl
}


\institute{Sebastian Baltes \at
		University of Trier, Germany \\
		\email{research@sbaltes.com} 
		\and
		Lorik Dumani \at
		University of Trier, Germany \\
		\email{dumani@uni-trier.de} 
		\and
		Christoph Treude\at
		University of Adelaide, Australia \\
		\email{christoph.treude@adelaide.edu.au} 
		\and
		Stephan Diehl \at
		University of Trier, Germany \\
		\email{diehl@uni-trier.de} 
}

\date{Received: date / Accepted: date}

\maketitle

\begin{abstract}
Stack Overflow (SO) is the most popular question-and-answer website for software developers, providing a large amount of code snippets and free-form text on a wide variety of topics.
Like other software artifacts, questions and answers on SO evolve over time, for example when bugs in code snippets are fixed, code is updated to work with a more recent library version, or text surrounding a code snippet is edited for clarity.
To be able to analyze how content on SO evolves, we built \emph{SOTorrent}, an open dataset based on the official SO data dump.
\emph{SOTorrent} provides access to the version history of SO content at the level of whole posts and individual text or code blocks.
It connects SO posts to other platforms by aggregating URLs from text blocks and comments, and by collecting references from GitHub files to SO posts.
In this paper, we describe how we built \emph{SOTorrent}, and in particular how we evaluated 134 different string similarity metrics regarding their applicability for reconstructing the version history of text and code blocks.
Based on different analyses using the dataset, we present:
(1) insights into the evolution of SO posts, e.g., that post edits are usually small, happen soon after the initial creation of the post, and that code is rarely changed without also updating the surrounding text;
(2) a qualitative study investigating the close relationship between post edits and comments,
(3) a first analysis of code clones on SO together with an investigation of possible licensing risks.
Finally, since the initial presentation of the dataset, we improved the post block extraction and our predecessor matching strategy.

\end{abstract}

\section{Introduction}



Stack Overflow (SO) is the most popular question-and-answer website for software developers.
As of December 2017, its public data dump~\citep{StackExchangeInc2017b} lists over 38 million posts and over 8 million registered users.
Many answers contain code snippets together with explanations~\citep{YangHussainOthers2016}.
Similar to other software artifacts such as source code files and documentation~\citep{Lehman1980, ChapinHaleOthers2001, MensDemeyer2008, GodfreyGerman2008}, text and code snippets on SO evolve over time, e.g., when the SO community fixes bugs in code snippets, clarifies questions and answers, and updates documentation to match new API versions.
Since the inception of SO in 2008, a total of 13.9 million SO posts have been edited after their creation---19,708 of them more than ten times. 
While many SO posts contain code, the evolution of code snippets on SO differs from the evolution of entire software projects: Most snippets are relatively short (on average 12 lines, see Section~\ref{sec:analysis-evolution}) and many of them cannot compile without modification~\citep{YangHussainOthers2016}.
In addition, SO does not provide a version control or bug tracking system for post content, forcing users to rely on the commenting function or additional answers to voice concerns about a post.

Recent studies have shown that developers use SO snippets in their software projects, regardless of maintainability, security, and licensing implications~\citep{BaltesKieferOthers2017, AnMloukiOthers2017, YangMartinsOthers2017,  GharehyazieRayOthers2017, AbdalkareemShihabOthers2017, XiaBaoOthers2017, FischerBottingerOthers2017, AcarBackesOthers2016}.
Assuming that developers copy and paste snippets from SO without trying to thoroughly understand them, maintenance issues arise.
For instance, it may later be more difficult for developers to refactor or debug code that they did not write themselves.
Moreover, if no link to the SO post is added to the copied code, it is not possible to check the SO thread for a corrected or improved solution in case problems occur. 
The same holds for code clones within Stack Overflow, which themselves may have been copied from external sources into Stack Overflow posts.
These complicated relationships may not only lead to issues affecting the maintainability of the code snippets on Stack Overflow or their copies in software projects or documentation resources, but also to licensing issues when people do not adhere to the license of the original content.

The SO data dump keeps track of different versions of entire posts, but does not contain information about differences between versions at a more fine-grained level.
In particular, it is not trivial to extract different versions of the same code snippet from the history of a post to analyze its evolution or compare code snippets between posts.
To address these challenges, we have created the open dataset \emph{SOTorrent}~\citep{BaltesDumaniOthers2018}, which enables researchers to analyze the version history of SO posts at the level of whole posts and individual post blocks, and their relation to corresponding source code in GitHub repositories.
Beside describing how we created that dataset, we use it to answer four research questions about the evolution of SO posts:

\begin{itemize}[labelindent=\parindent, labelwidth=\widthof{\textbf{RQ1:}}, label=\textbf{RQ1:}, leftmargin=*, align=parleft, parsep=0pt, partopsep=0pt, topsep=1ex, noitemsep]
\item[\textbf{RQ1:}] How do Stack Overflow posts evolve?
\item[\textbf{RQ2:}] Which posts get edited?
\item[\textbf{RQ3:}] Which edit and communication patterns exist?
\item[\textbf{RQ4:}] What are the implications of code clones on Stack Overflow?
\end{itemize}

While answering the first two questions will further our understanding of the phenomenon of SO post evolution, the third question aims at finding a connection between post edits and other events on the SO platform.
The fourth question transfers a well-known software engineering problem affecting the maintainability of software~\citep{JuergensDeissenboeckOthers2009, ThummalapentaCeruloOthers2010} to code snippets on SO.

The first two research question have already been covered in our previous conference paper~\citep{BaltesDumaniOthers2018}, but we added a comparison of the evolution of questions and answers (end of Section~\ref{sec:analysis-evolution}).
The results from our previous work motivated a further investigation of the relationship between post edits and comments, which yielded the edit and communication patterns we present in this paper (Section~\ref{sec:edits-comments}).
The fourth research question demonstrates how the \emph{SOTorrent} dataset can be used to analyze code snippets on SO, but also points to an underexplored phenomenon: code clones on Stack Overflow (Section~\ref{sec:code-clones}).
Finally, we analyzed all false positive and false negative results that our previous post history reconstruction approach yielded for our test dataset (Section~\ref{sec:analysis-f}) and revised the matching strategy according to our observations (Section~\ref{sec:revised-strategy}).
In the end, we were able to resolve the matching issues for 90.4\% of the affected posts.

We found that SO posts grow over time in terms of their number of text and code blocks, but the size of the individual blocks is relatively stable.
Many edits ($44.1\%$) just modify a single line of text or code, but only in $6.1\%$ of the cases are code blocks changed without also changing text content; post edits usually happen shortly after the creation of the post.
Our research suggests that comments and post edits are closely related: Some comments might trigger edits, others might be made in response to the edits.
We investigated 213 edit and comment events from 58 different SO posts and describe six edit and communication patterns that we observed.
Regarding the code clones, we used \emph{SOTorrent} to detect them, qualitatively investigated the source of 50 frequently copied snippets, and started a discussion in the SO community about possible implications and strategies to handle code clones.

\section{The SOTorrent Dataset}

\begin{figure}
\centering
\includegraphics[width=0.95\columnwidth,  trim=0.0in 0.0in 0.0in 0.0in]{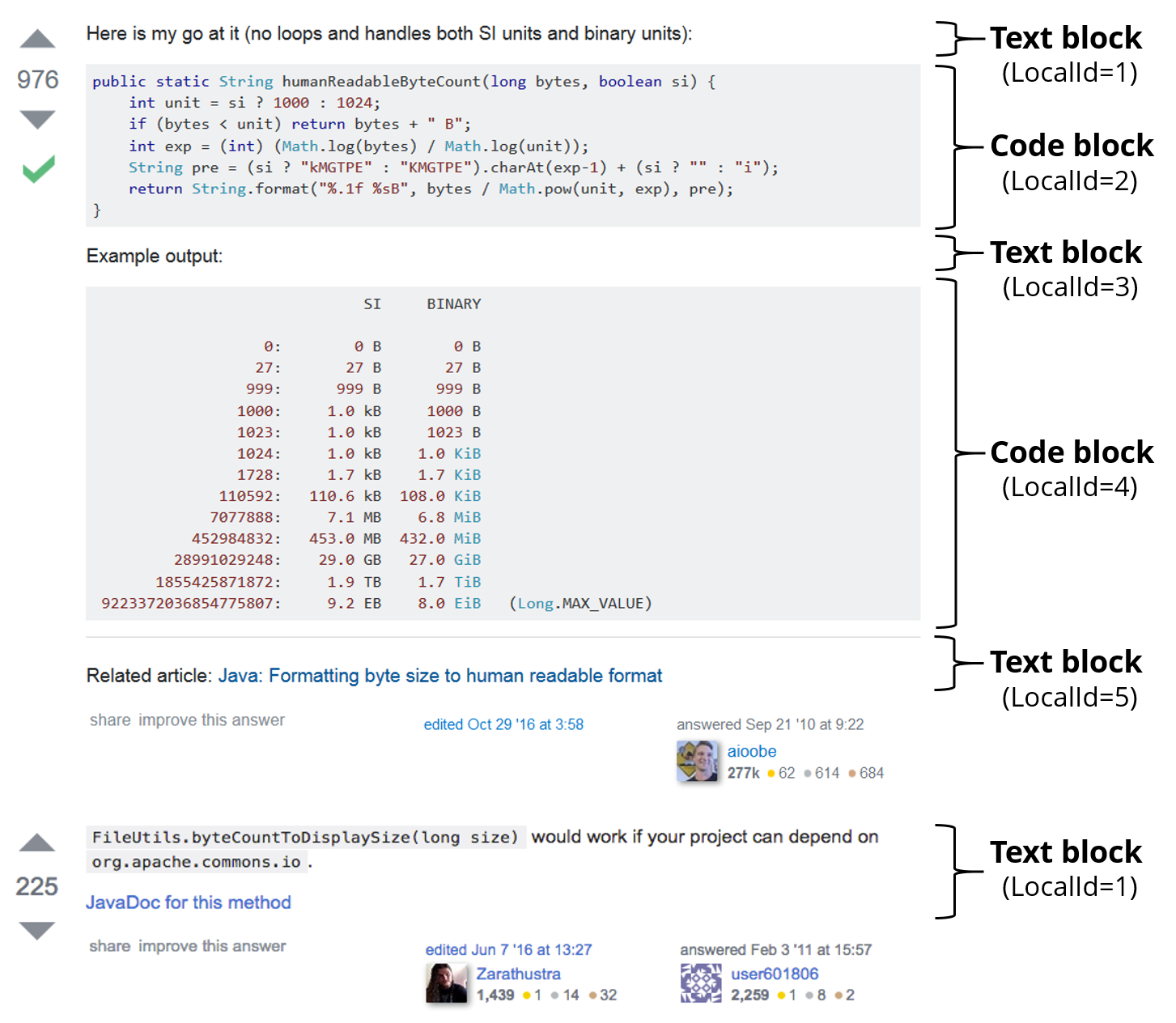} 
\caption{Exemplary Stack Overflow answers with code blocks (top, \href{https://stackoverflow.com/a/3758880}{3758880}) and with inline code (bottom, \href{https://stackoverflow.com/a/4888400}{4888400}). The \texttt{LocalId} represents the position in the post.}
\label{fig:so-postblocks-example}
\end{figure}

To answer our research questions, and to support other researchers in answering similar questions, we build \emph{SOTorrent}, an open dataset based on data from the official SO data dump~\citep{StackExchangeInc2017b} and the Google BigQuery GitHub (GH) dataset~\citep{GoogleCloudPlatform2018}.
\emph{SOTorrent} provides access to the version history of SO content at the level of whole posts and individual post blocks.
A post block can either be a text or a code block, depending on how the author formatted the content (see Figure~\ref{fig:so-postblocks-example} for an example).
Beside providing access to the version history, the dataset links SO posts to external resources in two ways: (1) by extracting linked URLs from text blocks and comments on SO and (2) by providing a table with links to SO posts found in the source code of GitHub projects.
This table can be used to connect \emph{SOTorrent} and GH datasets such as \emph{GHTorrent}~\citep{Gousios2013}.
Our dataset is available on Zenodo as a database dump~\citep{BaltesDumani2018}, including instructions on how to import the dataset, and as a public BigQuery dataset.\footnote{\url{https://bigquery.cloud.google.com/dataset/sotorrent-org:2018_09_23}}
We also published the source code of the software that we used to build~\citep{BaltesDumani2018c, Baltes2018b} and analyze~\citep{Baltes2018d, Baltes2018e} \emph{SOTorrent}.

\begin{figure}
\centering
\includegraphics[width=0.8\columnwidth,  trim=0.0in 0.0in 0.0in 0.0in]{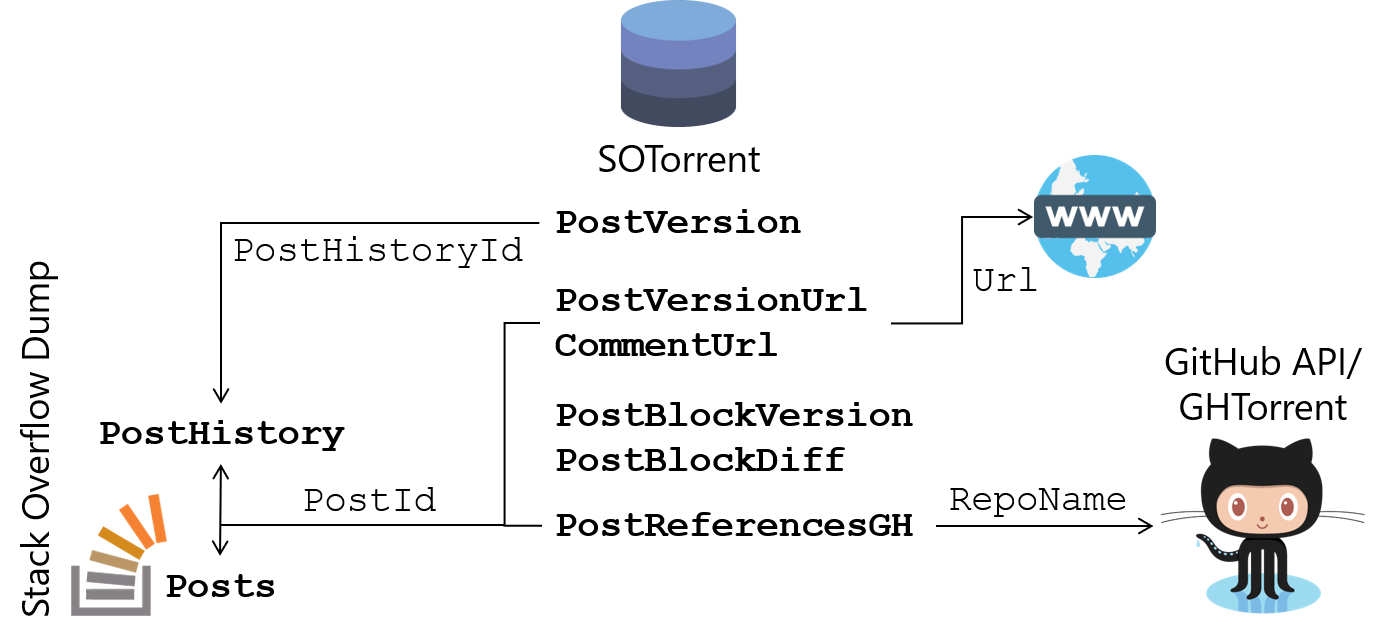}
\caption{Connection of \emph{SOTorrent} tables to other resources.}
\label{fig:linking}
\end{figure}

\emph{SOTorrent} release 2018-08-28, for example, contains the version history of all 40,606,950 questions and answers in the official SO data dump published June 5, 2018~\citep{StackExchangeInc2017b}.
It contains 63,914,798 post versions, 122,673,430 text block versions, and 77,578,494 code block versions, ranging from the creation of the first post on July 31, 2008 until the last edit on June 3, 2018.
We extracted links to 11,775,659 distinct URLs from 20,518,181 different post block versions and 4,104,869 distinct URLs from 6,856,777 different comments.
Moreover, we identified 6,035,737 links to SO posts in 436,615 public GH repositories. 
Our project website\footnote{\url{http://sotorrent.org}} lists all dataset versions and contains more information on the database layout, including the complete database schema.
In the following sections, we provide information about \emph{SOTorrent}'s data storage and collection process, before we use the dataset to answer our research questions.

\section{Database Schema}
\label{sec:database-schema}

\begin{figure}
\centering
\includegraphics[width=1.0\textwidth,  trim=2.7in 3.0in 2.7in 3.0in]{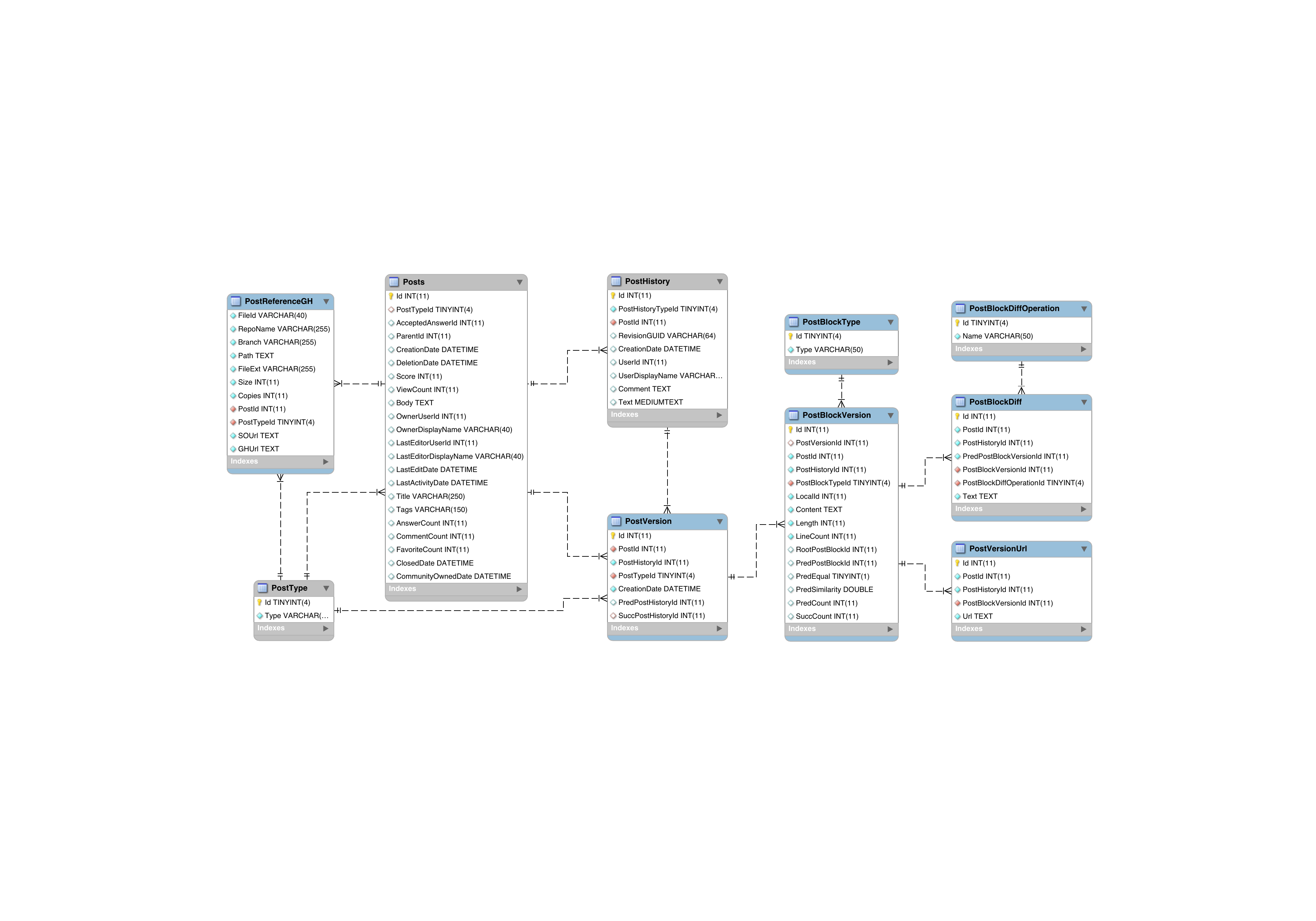} 
\caption{Database schema of \emph{SOTorrent} release \emph{2018-02-16}: The tables from the official SO dump~\citep{StackExchangeCommunityWiki20180227} are marked gray, the additional tables are marked blue. Not all tables from the official SO dump and not all foreign key constraints are shown. The most recent version of the database schema is always available on the \emph{SOTorrent} \href{http://sotorrent.org}{project page}.}
\label{fig:schema}
\end{figure}

\emph{SOTorrent} contains all tables from the official Stack Overflow data dump (see database schema in Figure~\ref{fig:schema}). 
Figure~\ref{fig:linking} visualizes how the \emph{SOTorrent} tables are connected to the SO dump, external resources on the web, and projects on GitHub.
The official data dump only provides the version history at the level of whole posts as Markdown-formatted text.
To analyze how individual text or code blocks evolve, we needed to extract individual blocks from that content.
This extraction also enabled us to collect links to external sources from the identified text blocks.
In the SO dump, one version of a post corresponds to one row in the table \texttt{PostHistory}.
However, that table does not only document changes to the content of a post, but also changes to metadata such as tags or title.
Since our goal was to analyze the evolution of SO posts at the level of whole posts and individual post blocks, we had to filter and process the available data.
First, we selected edits that changed the content of a SO post, identified by their \texttt{PostHistoryTypeId}~\citep{StackExchangeCommunityWiki20180227} ($2$: \textit{Initial Body}, $5$: \textit{Edit Body}, $8$: \textit{Rollback Body}).
We linked each filtered version to its predecessor and successor and stored it in table \texttt{PostVersion}.

The content of a post version is available as Markdown-formatted text.
We split the content of each version into text and code blocks and extracted the URLs from all text blocks using a regular expression (table \texttt{PostVersionUrl}).
We also extracted the URLs from all comments in the SO data dump (table \texttt{CommentUrl}).
Beside the extracted URLs, those tables provide information about the link type (e.g., bare, Markdown, or HTML), link position (top, middle, or end of post/comment), and certain URL components such as the root domain, query string, or fragment identifier (if present).

To reconstruct the version history of individual post blocks, we established a linear predecessor relationship between the post block versions (table \texttt{PostBlockVersion}) using a string similarity metric that we selected after a thorough evaluation (see Section~\ref{sec:metrics-evaluation}).
For each post block version, we computed the line-based difference to its predecessor, which is available in table \texttt{PostBlockDiff}.
We also extracted the version history of question titles from table \texttt{PostHistory}.
Table \texttt{TitleVersion} links all title versions to their predecessors and successors and further provides the corresponding Levenshtein distances (columns \texttt{PredEditDistance} and \texttt{SuccEditDistance}).

One row in table \texttt{PostReferenceGH} represents one link from a file in a public GH repository to a post on SO.
To extract those references, we utilized Google BigQuery, which allows to execute SQL queries on various public datasets, including a dataset with all files in the default branch of GH projects~\citep{GoogleCloudPlatform2018}.
To find references to SO, we again applied a regular expression and mapped all extracted URLs to their corresponding sharing link (ending with \verb+/q/<id>+ for questions and \verb+/a/<id>+ for answers), storing that link together with information about the file and the repository in which the link was found in table \texttt{PostReferenceGH}.
We ignored other links referring to, e.g., users or tags on SO.

\section{Post Block Extraction}
\label{sec:postblock-extraction}

Our goal was to analyze the evolution of individual text and code blocks, for example to trace changes to particular code snippets, to find code clones on SO, or to identify bug fixes for code on SO.
Moreover, the differentiation between the two post block types allowed us to extract links to external resources only from text blocks, not from code blocks.
The latter may, for example, contain XML namespace links or links to stylesheet files, which we do not consider to be external sources of the post.
The first step towards reconstructing the version history of individual post blocks is their extraction from the Markdown-formatted text that SO uses for the content of posts.
In our notion, a code block is not a short inline code fragment embedded into a text block (see Figure~\ref{fig:so-postblocks-example} for an example), but a continuous code snippet.
We consider inline-code to be part of the surrounding text block.
According to SO's Markdown specification~\citep{StackExchangeInc2018}, code blocks are indented by four spaces and inline code is framed by backtick characters.
However, as we found during our research, users are free to use other Markdown specifications or HTML tags, which are not officially supported, but correctly parsed and displayed on the SO website.
We iteratively tested and refined our post block extraction approach using a random sample of over 100,000 SO posts ($s_\textit{large}$).
We ran the extraction, randomly checked the extracted posts blocks, and added a new test case if the result differed from the rendering on the SO website (class \texttt{PostVersionHistoryTest}~\citep{BaltesDumani2018c}).
We then updated the extraction such that all test cases passed and re-ran the extraction on the test data.
The final version of our post block extraction method was able to detect various notations that SO authors used to mark code blocks, including SO Markdown (indented by 4 spaces), code fencing Markdown (enclosed by three backticks), SO stack snippets (enclosed by \texttt{<!--begin/end snippet-->}), stack snippet language tags (prepended by \texttt{<!--language:...-->}), HTML code tags (enclosed by \texttt{<pre><code>}), and HTML script tags (enclosed by \texttt{<script>}).

\section{Post Block Matching}

After successfully extracting the post blocks from a post version, we had to map the extracted post blocks to their predecessors in the previous post version to reconstruct their version history.
Since this mapping had to work for text and code content, the latter in various programming languages, we decided to utilize syntax-based similarity metrics.
We implemented 134 different string similarity metrics (see Section~\ref{sec:similarity-metrics}), which we evaluated regarding their correctness and performance using the manually validated version history of 600 SO posts (see Sections~\ref{sec:ground-truth} and \ref{sec:metrics-evaluation}).
In case of multiple matches, we had to choose between different predecessor candidates.
Thus, we developed a matching strategy that considers the location and context of a post block (see Section~\ref{sec:matching-strategy}).

\subsection{Similarity Metrics}
\label{sec:similarity-metrics}

\begin{table}
\setlength\tabcolsep{4pt}
\caption{Overview of all implemented base similarity metrics ($n=32$).}
\small
\begin{tabular}{lll}
\hline\noalign{\smallskip}
\textbf{Type} & \multicolumn{2}{l}{\textbf{Metric}} \\
\noalign{\smallskip}\hline\noalign{\smallskip}
\multirow{2}{*}{edit} & levenshtein & damerauLevenshtein \\
& longestCommonSubsequence (LCS) & optimalAlignment (OA) \\
\noalign{\smallskip}\hline\noalign{\smallskip}
\multirow{2}{*}{set} & nGram\{Jaccard$|$Dice$|$Overlap\} & nShingle\{Jaccard$|$Dice$|$Overlap\} \\
& token\{Jaccard$|$Dice$|$Overlap\} & \\
\noalign{\smallskip}\hline\noalign{\smallskip}
\multirow{3}{*}{profile} & cosineNGram\{Bool$|$TF$|$NormalizedTF\} & manhattanNGram  \\
& cosineNShingle\{Bool$|$TF$|$NormalizedTF\} & manhattanNShingle \\
& cosineToken\{Bool$|$TF$|$NormalizedTF\}  & manhattanToken \\
\noalign{\smallskip}\hline\noalign{\smallskip}
\multirow{2}{*}{fingerprint} & \multicolumn{2}{l}{\multirow{2}{*}{winnowingNGram\{Jaccard$|$Dice$|$Overlap$|$LCS$|$OA\}}} \\
&  & \\
\noalign{\smallskip}\hline\noalign{\smallskip}
\multirow{1}{*}{equal} & equal & tokenEqual \\
\noalign{\smallskip}\hline
\end{tabular}
\label{tab:similarity-metrics-1}
\end{table}

\begin{table}
\setlength\tabcolsep{4pt}
\caption{Overview of all evaluated variants of the implemented similarity metrics ($n=134$).}
\small
\begin{tabular}{lll}
\hline\noalign{\smallskip}
\textbf{Type} & \textbf{Variants} \\
\noalign{\smallskip}\hline\noalign{\smallskip}
\multirow{2}{*}{edit} & \multirow{2}{*}{\shortstack[l]{with/without normalization}}\\
\\
\noalign{\smallskip}\hline\noalign{\smallskip}
\multirow{2}{*}{set} & \multirow{2}{*}{\shortstack[l]{$n\text{Gram} : n \in \{2,3,4,5\}$, $n\text{Shingle} : n \in \{2,3\}$\\ with/without normalization, padding (nGram)}}\\
\\
\noalign{\smallskip}\hline\noalign{\smallskip}
\multirow{3}{*}{profile} & \multirow{3}{*}{\shortstack[l]{$n\text{Gram} : n \in \{2,3,4,5\}$, $n\text{Shingle} : n \in \{2,3\}$\\ with normalization (both) and without (cosine)}}\\
\\
\\
\noalign{\smallskip}\hline\noalign{\smallskip}
\multirow{2}{*}{fingerprint} & \multirow{2}{*}{\shortstack[l]{$n\text{Gram} : n \in  \{2,3,4,5\}$, \\ with/without normalization}}\\
\\
\noalign{\smallskip}\hline\noalign{\smallskip}
\multirow{1}{*}{equal} & \multirow{1}{*}{\shortstack[l]{with/without normalization}}\\
\noalign{\smallskip}\hline
\end{tabular}
\label{tab:similarity-metrics-2}
\end{table}

A similarity metric maps two input strings to a value in $[0, 1]$, where $0$ corresponds to  inequality and $1$ corresponds to equality.
We implemented five different types of similarity metrics: \emph{edit-based metrics} (e.g., Levenshtein), \emph{set-based metrics} (e.g., n-grams with Jaccard coefficient), \emph{profile-based metrics} (e.g, cosine similarity), \emph{fingerprint-based metrics} (Winnowing), and \emph{equality-based metrics}, which served as a baseline in the metrics evaluation (see Section~\ref{sec:metrics-evaluation}).
Our Java implementation of all metrics is available on GitHub~\citep{BaltesDumani2018d}.
Tables~\ref{tab:similarity-metrics-1} and \ref{tab:similarity-metrics-2} show all metrics that we implemented and evaluated.

The \emph{edit-based metrics} define the similarity of two strings based on the number of edit operations needed to transform one string into the other.
Optimal string alignment (OA) allows the two operations `insertion of one character' and `deletion of one character'.
The Levenshtein distance further allows `substitution of one character'.
The Damerau-Levenshtein distance is similar to Levenshtein, but additionally allows the operation `swap two neighboring characters'.
The longest common subsequence (LCS) of two strings is the longest sequence of characters (order irrelevant) that can be found in both strings.
It can be interpreted as a variant of Damerau-Levenshtein with the additional restriction that each character can only be modified once (e.g., swapping two characters and then replacing one of them is not possible).
To derive a similarity metric from the number of edit operations and the longest common subsequence, we used the following approaches:

\begin{definition}[Edit/LCS Similarity]
Let $S_1$, $S_2$ be two strings, $d$ be the edit distance and $LCS$ be the longest common subsequence between the two strings: $(S_1, S_2)  \to \mathbb{R}_{0}^{+}$. The edit-based and the LCS-based similarity functions $sim \colon (S_1, S_2) \to [0, 1]$ are then defined as
\[ sim_\text{edit}(S_1, S_2) = \frac{max(|S_1|, |S_2|) - d(S_1, S_2)}{max(|S_1|, |S_2|)} \]
\[ sim_\text{lcs}(S_1, S_2) = \frac{LCS(S_1, S_2)}{max(|S_1|, |S_2|)} \]
\end{definition}

The \emph{profile-based metrics} consider each distinct token, n-gram, or n-shingle in the two input strings as one dimension of a vector space.
Tokens can be extracted from a string by a tokenization with whitespaces as delimiter, $n$-grams split the string in sequences of $n$ consecutive characters, $n$-shingles split the string in sequences of $n$ consecutive words or tokens.
One input string is then characterized as one vector in the vector space.
In the simplest form (bool), the values of the dimensions can either be 1 (token, n-gram, or n-shingle present in the string) or 0 (not present).
Alternatively, one can consider the number of occurrences of each token, n-gram, or n-shingle as the value of the dimensions (term frequency).
We also considered the BM15 weighting scheme ($k=1.5$)~\citep{ManningRaghavanOthers2008}, which intends to lower the effect of very frequent terms skewing the comparison.
The similarity of the two strings is then defined as the cosine or Manhattan distance between the two vectors that have been derived from the strings using one of the three approaches described above.

For the \emph{set-based metrics}, we considered all distinct tokens, n-grams and n-shingles in the strings as elements of sets. We used three coefficients to compare the resulting sets:

\begin{definition}[Similarity Coefficients]
Let $S_1$, $S_2$ be sets of tokens, n-grams, or n-shingles.
\begin{align*}
Jaccard(S_1, S_2) &= \frac{|S_1 \cap S_2|}{|S_1 \cup S_2|} \hspace{2.2em} Dice(S_1, S_2) = \frac{2 \cdot |S_1 \cap S_2|}{|S_1| + |S_2|} \\
Overlap(S_1, S_2) &= \frac{|S_1 \cap S_2|}{min(|S_1|, |S_2|)} & \\
\end{align*}
\end{definition}

\begin{figure}
\centering
\includegraphics[width=1\columnwidth,  trim=0.0in 0.95in 0.0in 0.0in]{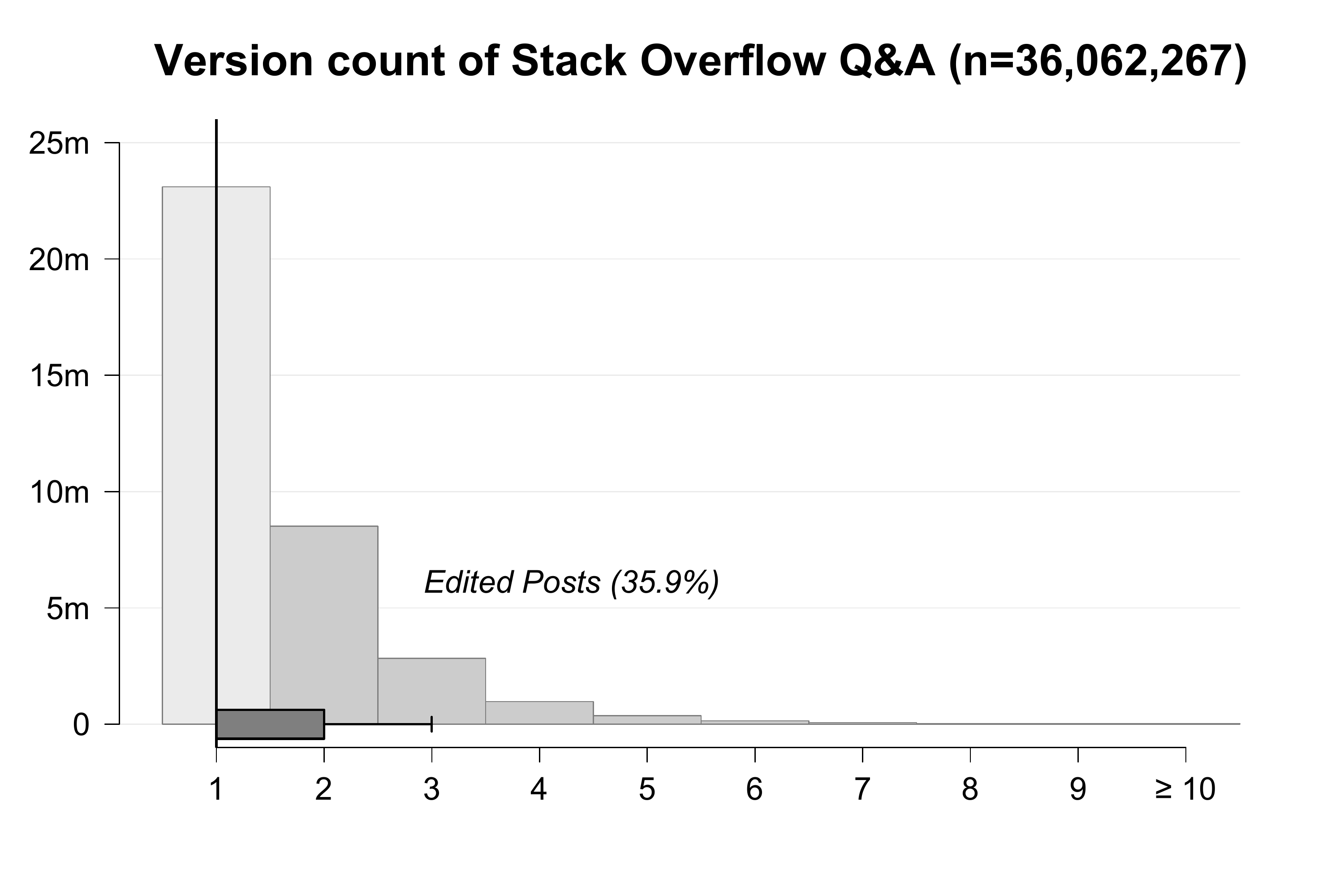} 
\caption{Histogram and boxplot showing the number of Stack Overflow questions and answers with a certain version count (PostHistoryTypeIds 2, 5, 8); based on the SO data dump 2017-06-12; vertical line is median.}
\label{fig:version-count}
\end{figure}

The \emph{fingerprint-based metrics} apply a hash function to substrings of the input strings and then use the computed hash values to determine the similarity.
The Winnowing algorithm is one approach to calculate and compare the fingerprints of two strings~\citep{SchleimerWilkersonOthers2003, DuricGasevic2013}.
Winnowing is often used for plagiarism detection, e.g., in the source code comparison software \emph{MOSS}~\citep{BurrowsTahaghoghiOthers2007, MartinsFonteOthers2014, LancasterCulwin2004}.
We implemented different variants of the algorithm described by Schleimer et al.~\citep{SchleimerWilkersonOthers2003}, e.g., using different n-grams sizes and different approaches to compare the fingerprints.

We implemented each metric in different variations.
In the variants with normalized input strings, we used different approaches for different metric types:
For the edit metrics, we unified the whitespace characters, i.e., reduced them to a single space, and converted all characters to lower case.
For the n-gram metrics, we converted all characters to lower case, removed all whitespace, and removed some special characters (\texttt{\{\};}) (see Section~\ref{sec:analysis-f} for the characters we later added to this set).
For the shingle metrics, we again converted all characters to lower case, unified the whitespace characters, and removed all non-word characters (\texttt{[\^{}a-zA-Z\_0-9]}). 
We used common n-gram and shingle sizes~\citep{BurrowsTahaghoghiOthers2007} and also implemented an optional n-gram padding that emphasizes the beginning and the end of the input strings.
All these variations lead to a total number of 134 different similarity metrics.

\begin{figure}
\centering
\includegraphics[width=1\textwidth,  trim=0.0in 0.0in 0.0in 0.0in]{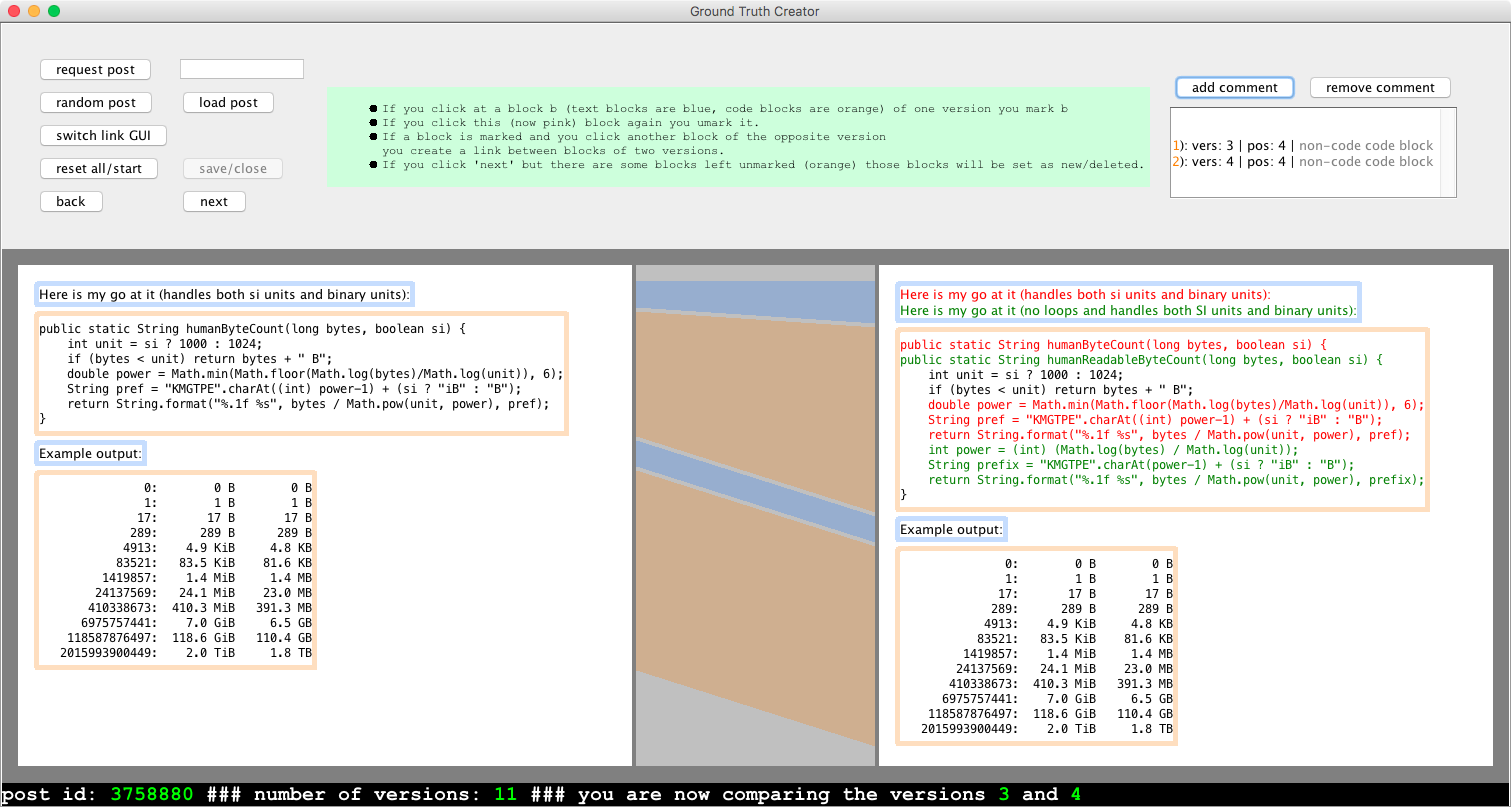} 
\caption{App developed to create ground truth for similarity metric evaluation.}
\label{fig:gt-app}
\end{figure}

\begin{figure}
\centering
\includegraphics[width=1\columnwidth,  trim=0.0in 0.0in 0.0in 0.0in, frame]{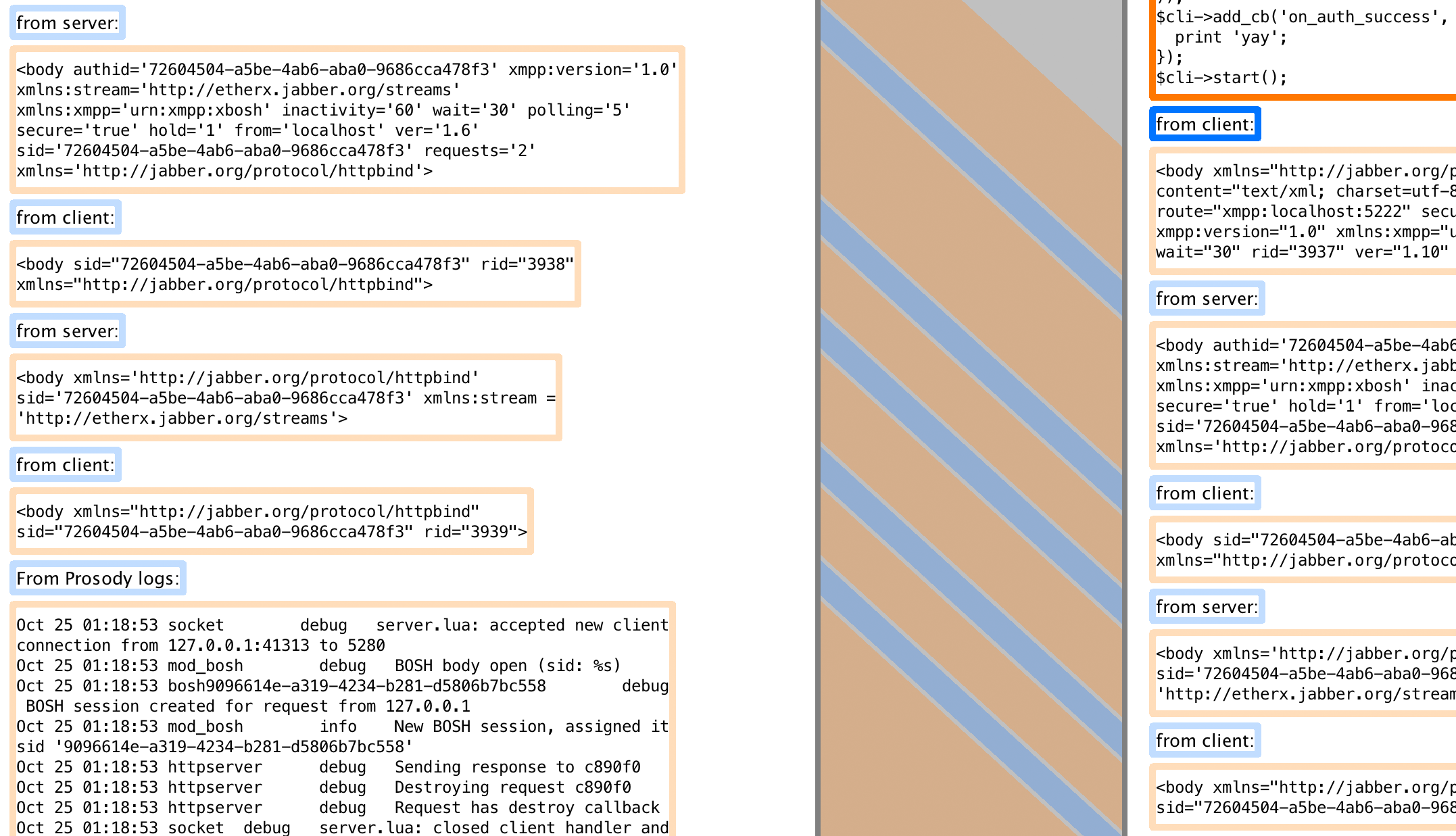} 
\caption{Post with multiple equal predecessors (\href{https://stackoverflow.com/posts/13064858/revisions}{13064858}).}
\label{fig:multiple-equal-predecessors}
\end{figure}

\subsection{Ground Truth}
\label{sec:ground-truth}

To evaluate the correctness of the post block mappings retrieved using different string similarity metrics, we created a set of 600 manually validated post version histories.
Figure~\ref{fig:gt-app} shows a screenshot of the tool we developed to create those manually validated histories (available on GitHub~\citep{DumaniBaltes2017}).
It visualizes a post version (right) and its predecessor (left).
Post blocks with equal content and type that are unique in the two versions are automatically connected.
For the other post blocks, the user has to choose a match by clicking on a post block of the same type in each version; the tool then visualizes the line-based difference between the connected blocks.
It is also possible to add comments for individual post blocks, e.g., in case the user is not confident in his or her mapping, or in case the post block extraction failed. 

We drew four different samples from the SO data dump released June 12, 2017.
The first sample with 200 posts ($s_\textit{rand}$) was randomly drawn from all SO questions and answers with at least two versions (otherwise no mapping is needed).
Since there are many posts with only two versions (see Figure~\ref{fig:version-count}), we decided to draw another sample of 200 posts from SO questions and answers with at least seven versions (99\% quantile).
We denote this sample $s_\textit{rand+}$.
As the initial focus of our research was on Java, we also drew a sample with 200 Java posts ($s_\textit{java}$) from all SO questions tagged with \texttt{<java>} or \texttt{<android>}, and the corresponding answers.
The last sample ($s_\textit{mult}$), which contains 100 posts with multiple possible predecessors, was not used to evaluate the metrics, but to evaluate our matching strategy (see Section~\ref{sec:matching-strategy}).
In this sample, we included posts which had at least two possible matches (two post blocks of the same type with identical content) in two adjacent versions.

The validated version histories of the samples were created by a graduate student, and then later discussed with two of the authors.
The student was introduced to the app and told to comment on all post blocks where he is not sure about the mapping. 
Together, we looked at all post blocks with comments indicating an unclear mapping ($n=38$) and tried to find a mapping we all agreed on.
If that was not possible, we moved the post to a new sample $s_\textit{unclear}$, which we separately analyzed.
After discussing all 38 posts, $s_\textit{unclear}$ contained 17 posts (4 from $s_\textit{rand}$, 8 from $s_\textit{rand+}$, and 5 from $s_\textit{java}$).
All samples are available on Zenodo~\citep{BaltesDumaniOthers2017}.

\subsection{Matching Strategy}
\label{sec:matching-strategy}

Our goal was to establish a linear predecessor relationship for all post block versions, thus each post block version can only have one predecessor.
The reason for this decision was the fact that we rarely observed splits and merges in the post version histories we manually analyzed.
Moreover, even if multiple predecessors have equal or similar content, usually only one of them is the actual predecessor (see Figure~\ref{fig:multiple-equal-predecessors} for an example).
To correctly choose the predecessor from different candidates, we had to develop a matching strategy for post block versions, which we present in this section.
In the database, we not only store the matched predecessor, but also the number of possible predecessors and successors, to be later able to identify post version histories that could contain splits or merges. 
For our analysis (see Section~\ref{sec:analysis}), we consider \textit{post block lifespans}, i.e., chains of connected post block versions that are predecessors of each other.
Those lifespans can be easily retrieved from the database, because each post block version has a \texttt{RootPostBlockVersionId}, which is the id of the first post block version in the chain.
Those chains can likewise be retrieved using the columns \texttt{RootPostHistoryId} and \texttt{RootLocalId}, which also uniquely identify the first post block version in a post block lifespan.

As mentioned above, we utilized a dedicated sample $s_\textit{mult}$ to evaluate how well our matching strategy can handle posts with multiple possible connections.
In case of differences between the ground truth and the results of our approach, we wrote unit tests replicating the issue and then updated the strategy until all unit tests passed.
We further used the sample $s_\textit{large}$ to test the strategy's scalability.
To be able to describe our matching strategy, we define our notation for post versions, post block versions, and possible predecessors:

\begin{definition}[Post Version]
Let $p$ be a post with $n$ versions.
Then $\,p_{i}$ denotes one post version and $|p_i|$ denotes the number of post blocks in $p_i$ for $i \in \{1 \ldots n\}$.
\end{definition}

\begin{definition}[Post Block Version]
Let $p_i$ be one post version and $\tau \in \{\text{text},~\text{code}\}$ be a post block type.
Then $\,b_{(i, l)}^{\tau}$ denotes one post block of type $\tau$ with local id $\,l$ for $\,l \in \{1 \ldots |p_i|\}$.
The function $id^\tau \colon p_i \to \{ 1 \le l \le |p_i| \}$ maps a post version to the local ids of the post blocks of type $\tau$ in that version.
\end{definition}

\begin{definition}[Possible Predecessors]
Let $b_{(i-1, l)}^{\tau}$, $b_{(i, j)}^{\tau}$ be post blocks of the same type in subsequent post versions,
\[ equal(b_{(i-1, l)}^{\tau} , b_{(i, j)}^{\tau}) \to \{ \text{true}, \text{false} \} \]
be a function that tests if the post blocks' contents are equal, and
\[ sim^\tau(b_{(i-1, l)}^{\tau} , b_{(i, j)}^{\tau}) \to [0,1] \]
be the similarity of the two post blocks' contents according to the similarity metric $sim^\tau$.
Let $\vartheta^\tau \in [0,1]$ be a threshold for $sim^\tau$.
Then, we define the set of equal predecessors as
\begin{align*}
PredEqual(b_{(i, j)}^{\tau})=\{ \beta^{\tau}_{(i-1, k)}~|~&equal(\beta^{\tau}_{(i-1, k)}, b_{(i, j)}^{\tau}) = \text{true},\\
& k \in id^\tau(p_{i-1}),~j \in id^\tau(p_{i}) \}
\end{align*}
We define the maximum predecessor similarity as
\begin{align*}
maxSim^\tau = max(\{ sim^\tau(&\beta^{\tau}_{(i-1, k)} , b_{(i, j)}^{\tau})~|~sim^\tau \ge \vartheta^\tau, \\
& k \in id^\tau(p_{i-1}),~j \in id^\tau(p_{i}) \} )
\end{align*}
In case no predecessor with a similarity above the threshold exists, we define $maxSim^\tau(\emptyset) = 0$. 
We define the set of matched predecessors as
\begin{align*}
PredMatched(b_{(i, j)}^{\tau})=\{ \beta^{\tau}_{(i-1, k)}~|~sim^\tau(&\beta^{\tau}_{(i-1, k)} , b_{(i, j)}^{\tau}) = maxSim^\tau, \\
& k \in id^\tau(p_{i-1}),~j \in id^\tau(p_{i}) \}
\end{align*}
Finally, we define the set of possible predecessors as
\[ Pred(b_{(i, j)}^{\tau}) =
\begin{cases}
PredEqual(b_{(i, j)}^{\tau}), &\mbox{if } PredEqual(b_{(i, j)}^{\tau}) \ne \emptyset , \\
PredMatched(b_{(i, j)}^{\tau}), &\mbox{if } PredEqual(b_{(i, j)}^{\tau}) = \emptyset .
\end{cases} \]
The set of possible successors $Succ(b_{(i, j)}^{\tau})$ is defined analogously.
\end{definition}

As can be seen in the above definition, we need two different similarity metrics ($sim^\textit{text}$ and $sim^\textit{code}$) and two different similarity thresholds ($\vartheta^\textit{text}$ and $\vartheta^\textit{code}$).
We only compute the similarity if the content of the post blocks is not equal, because we want to be able to distinguish equal post block versions from post block versions with similarity $1$ according to the metric.
Before we describe our matching strategy, we present two methods that we use in case of multiple possible predecessors.
Both methods iterate over all post blocks $b_{(i, j)}^{\tau}$ in a post version $p_{2 \le i \le n}$ that do not have a predecessor yet.
They follow different strategies for selecting a predecessor:

$setPredContext(p_i, BOTH)$ tries to select a predecessor using the post blocks before and after $b_{(i, j)}^{\tau}$, i.e., the blocks with local ids $j-1$ and $j+1$.
Please note that those blocks usually have a different post block type than $b_{(i, j)}^{\tau}$.
In case the predecessors of those neighboring blocks are already set and one post block $b_{(i-1, l)}^{\tau} \in Pred(b_{(i, j)}^{\tau})$ has the predecessors of those two post blocks as neighbors (local ids $l-1$ and $l+1$ in version $p_{i-1}$), the function sets $b_{(i-1, l)}^{\tau}$ as predecessor of $b_{(i, j)}^{\tau}$ and returns \textit{true}.
If no predecessor has been set, it returns \textit{false}.
In case of parameter \textit{ABOVE}, only the post block above (local id $j-1$) is taken into account; in case of parameter \textit{BELOW}, only the post block below (local id $j+1$) is taken into account.
Examples for posts that motivated this strategy are answer \href{https://stackoverflow.com/posts/32841902/revisions}{32841902} (mapping of version 2 to 1) and answer \href{https://stackoverflow.com/posts/37196630/revisions}{37196630} (mapping of version 2 to 1).

$setPredPosition(p_i)$ sets the post block $b_{(i-1, l)}^{\tau} \in Pred(b_{(i, j)}^{\tau})$ with $\Delta_\text{pos} = min(|l - j|)$, i.e., the post block with the local id closest to $j$, as predecessor of $b_{(i, j)}^{\tau}$.
If two possible predecessors have the same $\Delta_\text{pos}$, the method chooses the one with the smallest local id.
This approach is based on our observation that the order of post blocks rarely changes (see Section~\ref{sec:analysis-evolution}).
Examples for posts that motivated this strategy are question \href{https://stackoverflow.com/posts/18276636/revisions}{18276636} (mapping of version 2 to 1) and answer \href{https://stackoverflow.com/posts/2581754/revisions}{2581754} (mapping of version 3 to 2).

The complete matching strategy that selects (at most) one predecessor for each post block in a post version can be found as pseudo code in Algorithm~\ref{alg:matching-strategy}. 
The actual source code can be found in method \texttt{processVersionHistory} of class \texttt{PostVersionList} in the corresponding GitHub project~\citep{BaltesDumani2018c}.

\begin{algorithm}
\caption{Initial Matching Strategy}
\label{alg:matching-strategy}
\begin{algorithmic}
\ForAll{$p_{2 \le i \le n}$}
	\State \textit{// set predecessors where only one candidate exists}
	\ForAll{$b_{(i, 1 \le j \le |p_i|)}^{\tau}$}
		\If{$|Pred(b_{(i, j)}^{\tau})| = 1$}
			\State Let $pred$ be the equal or similar predecessor
			\If{$|Succ(pred)| = 1$}
				\State Set $pred$ as predecessor of $b_{(i, j)}^{\tau}$
				\State \textbf{continue}
			\EndIf
		\EndIf
	\EndFor
	\State \textit{// set predecessors using context}
	\State $predSet = \text{true}$
	\While{$predSet$}
		\State $predSet = setPredContext(p_i, BOTH)$
	\EndWhile
	\While{$predSet$}
		\State $predSet = setPredContext(p_i, BELOW)$
	\EndWhile
	\While{$predSet$}
		\State $predSet = setPredContext(p_i, ABOVE)$
	\EndWhile
	\State \textit{// set predecessors using position}
	\State $setPredPosition(p_i)$
\EndFor
\end{algorithmic}
\end{algorithm}

\subsection{Metrics Evaluation}
\label{sec:metrics-evaluation}

The matching strategy described above depends on the results of the similarity metrics $sim^\textit{text}$ and $sim^\textit{code}$ and the thresholds $\vartheta^\textit{text}$ and $\vartheta^\textit{code}$.
To select the best metrics for reconstructing the version history of post blocks, we evaluated all 134 metrics in different combinations with different thresholds using our ground truth samples $s_\textit{rand}$, $s_\textit{rand+}$, and $s_\textit{java}$.
Please note that the correctness of $sim^\textit{text}$ and $sim^\textit{code}$ cannot be evaluated independently, because the neighboring post blocks that $setPredContext$ takes into account usually have different types.
To assess the performance, we measured the runtime of the post history extraction for each configuration.
To assess the correctness of the extracted post block history, we regarded each metric configuration as a binary classifier that either assigns the predecessor of a post block version correctly or not (compared to the ground truth).
To calculate the number of true/false positives/negatives, we consider the set of \emph{predecessor connections}, i.e., all ($b_{(i-1, l)}^{\tau}$, $b_{(i, j)}^{\tau}$) from $\;p_{2 \le i \le n}$ that have been connected with a certain metric configuration.
We then compare those connections with the connections from the ground truth:

\begin{definition}[Metric Evaluation]
Let $\text{GT}^\tau$ be the set of predecessor connections of type $\tau$ in the ground truth, $\text{C}^\tau$ be the set of predecessor connections of type $\tau$ determined using a certain metric configuration,
and $\;n_\textit{pos}^\tau = \sum_{2 \le i \le n} |id^\tau(p_i)|\;$ be the number of possible predecessor connections of type $\tau$.
We define the number of true positives $\textit{tp}^\tau$, false positives $\textit{fp}^\tau$, true negatives $\textit{tn}^\tau$, and false negatives $\textit{fn}^\tau$ as:
\begin{align*}
\textit{tp}^\tau &= |\text{C} \cap \text{GT}| & \quad \textit{fp}^\tau &= |\text{C} \setminus \text{GT}|\\
\textit{tn}^\tau &= n_\textit{pos}^\tau - |\text{C} \cup \text{GT}| & \quad \textit{fn}^\tau &= |\text{GT} \setminus \text{C}|\\
\end{align*}
\end{definition}

After each comparison run, we ranked the configurations according to their Matthews correlation coefficient ($MCC$)~\citep{Matthews1975}, which takes $\textit{tp}^\tau$, $\textit{fp}^\tau$, $\textit{tn}^\tau$, and $\textit{fn}^\tau$ into account.
If two configurations had the same $MCC$ value, we ranked them according to their runtime.
$MCC$ is the preferred measure when evaluating binary classifiers~\citep{Chicco2017} and should be chosen over evaluation measures such as recall, precision, or F-measure~\citep{Powers2011}.
In our case, it correlates the connections from the ground truth and the connections set by a certain metric configuration.
The $MCC$ values are in range $[-1,1]$; a total disagreement is represented by $-1$, a perfect agreement by $1$.
The source code of the tool we used for the metrics evaluation is available on GitHub~\citep{BaltesDumani2018b}.

In the first comparison run, we configured $sim^\textit{text} = sim^\textit{code}$ and chose $\vartheta^\textit{\{text, code\}} \in \{0.0, 0.1, 0.2, \ldots, 1.0\}$.
This resulted in 1,474 different configurations.
The first run took about 24 hours on a regular desktop PC (Intel Core i7-7700, 64 GB RAM, 512 GB SSD). 

For the second run, we selected the metrics which, for a particular threshold, achieved an $MCC$ value in the 95\% quantile of all three samples either for text or for code blocks.
Some metrics cannot be applied to very short strings (e.g., if string length $<$ n-gram size).
For the final implementation, we wanted to have a backup metric that works for all input strings.
We filtered edit- and token-based metrics and selected the best candidates according to the criterion described above.
Finally, we selected 27 regular and 4 backup metrics for the second run.
We also added the \textit{equal} metric as a baseline.
We tested those metrics again with $sim^\textit{text} = sim^\textit{code}$, but this time we chose $\vartheta^\textit{\{text, code\}} \in \{0.0, 0.01, 0.02, \ldots, 1.0\}$
Thus, the second run tested 3,232 different configurations, which took about 20 hours.

As motivated above, the results of the text and code metrics depend on each other.
In the third and last run, we tested all combinations of the best (99\% quantile) text and code configurations together with the best backup configurations.
This was the first run with $sim^\textit{text} \ne sim^\textit{code}$ and with a backup metric for text and code blocks.
Those backup metrics were only used if the input strings were too short for the configured metrics.
The run, which took about 14 hours, tested all combinations of 13 text configurations, 3 text backup configurations, 15 code configurations, and 2 code backup configurations, resulting in 1,170 combinations in total.
For the final selection, we ranked the combinations according to the sum of their $MCC$ scores for text and code blocks.

The final configuration that we used to match post block predecessors for the \emph{SOTorrent} dataset was:

\begin{tabbing}
$sim^\textit{text}$ \hspace{0.7em} \= $=~\textit{manhattanFourGramNormalized}$\hspace{3em} \= ($\vartheta^\textit{text}\;$ \= $= 0.17$)\\
$sim^\textit{code}$ \> $=~\textit{winnowingFourGramDiceNormalized}$ \> ($\vartheta^\text{code}$ \> $= 0.23$)\\
$sim_\textit{backup}^\textit{text}$ \> $=~\textit{cosineTokenNormalizedTermFrequency}$ \> ($\vartheta^\textit{text}$ \> $= 0.36$)\\
$sim_\textit{backup}^\textit{code}$ \> $=~\textit{cosineTokenNormalizedTermFrequency}$ \> ($\vartheta^\textit{code}$ \> $=  0.26$)
\end{tabbing}

\begin{figure}
\centering
\includegraphics[width=0.9\columnwidth,  trim=0.0in 0.3in 0.2in 0.2in]{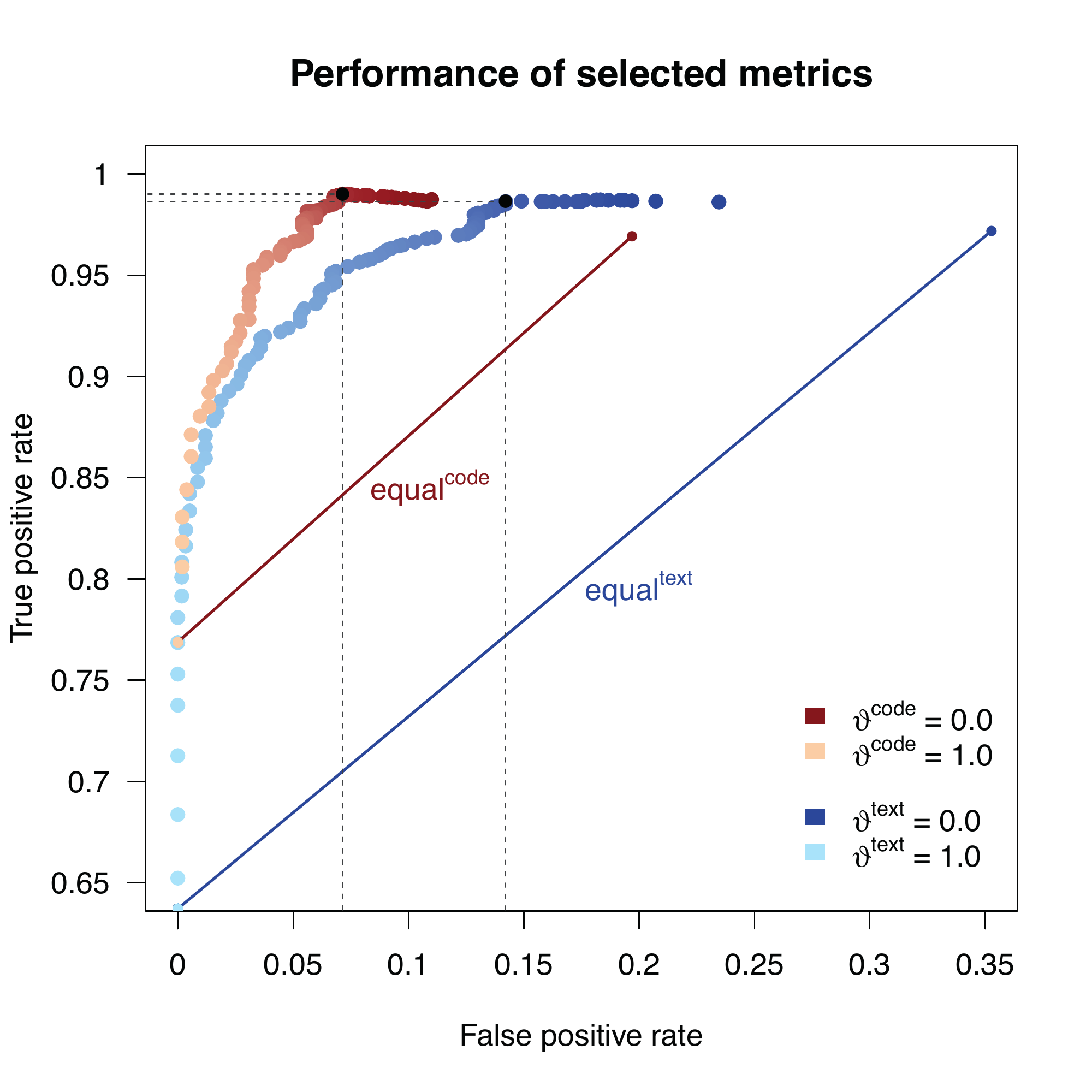} 
\caption{Performance of selected metrics: \textit{manhattanFourGramNormalized} for text (blue) and \textit{winnowingFourGramDiceNormalized} for code (red); selected thresholds: 0.17 for text and 0.23 for code (dotted lines).}
\label{fig:roc-excerpt}
\end{figure}

Figure~\ref{fig:roc-excerpt} shows the performance of the selected metrics for different thresholds with $sim^\textit{text}=sim^\textit{code}$, compared to the baseline metric \textit{equals}.
The final configuration achieved an $MCC$ value of $0.86$ for text (true positive rate $0.99$, false positive rate $0.14$) and $0.92$ for code (true positive rate $0.99$, false positive rate $0.07$).


%
%
%
%
%

\subsection{Analysis of False Positive and False Negative Predecessor Matches}
\label{sec:analysis-f}

\begin{figure}
\centering
\includegraphics[width=1\columnwidth,  trim=0.0in 0.3in 0.0in 0.0in]{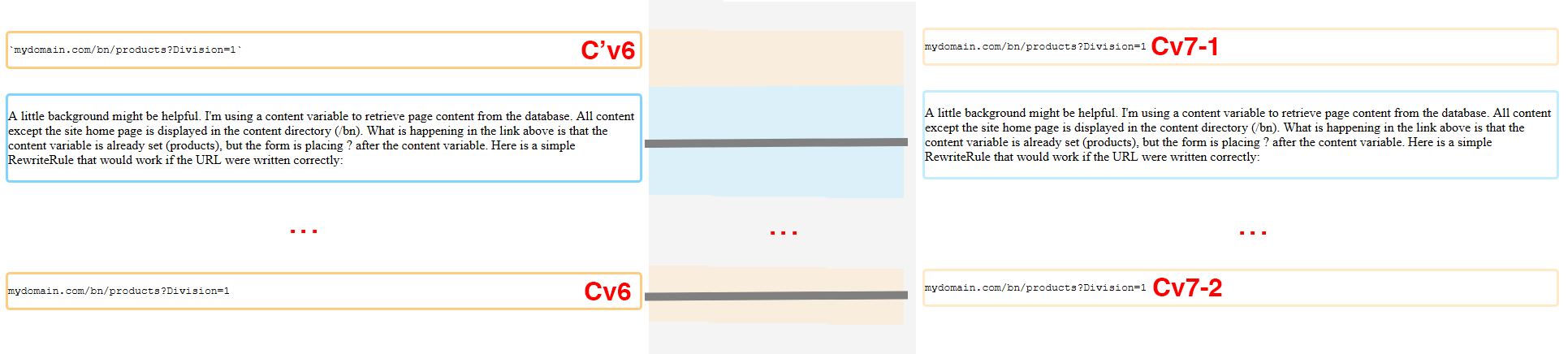} 
\caption{Issue with previous matching strategy in case the equal match is not available anymore (version 6 and 7 of question \href{https://stackoverflow.com/q/17158055}{17158055}): Orange/blue rectangles are connections in ground truth, lines are connections set by the previous matching strategy in combination with the selected default metrics; the connection between code blocks \textsc{C'v6} and \textsc{Cv7-1} is missing, because \textsc{Cv7-1} has an equal match in the previous version (\textsc{Cv6}), which is not available anymore; \textsc{C'v6} is very similar, but not equal to \textsc{Cv7-1}.}
\label{fig:matching-strategy-error}
\end{figure}

While the performance of our matching strategy together with the selected metrics was already good, we were eager to further reduce the number of false positives and negatives.
Therefore, we added a feature to our ground truth application that enabled us to display the difference between the ground truth and the mapping that our matching strategy with the default metrics produced (see Figure~\ref{fig:matching-strategy-error} for an example).
The source code of this revised application is available on GitHub~\citep{DumaniBaltes2018}.
We then systematically investigated all 31 posts with false positive or negative code block mappings, and then followed a similar approach as before to improve our matching strategy:
First, we decided whether an improved matching strategy could solve the observed matching problem and in case we agreed that it could, we created a test case reproducing the error.
This systematic approach lead to different improvements to the \emph{post block extraction}, the \emph{matching strategy}, and the \emph{default similarity metrics}, which we outline below.

In the end, we were able to solve the matching problem for 30 out of 31 posts.
In one post, the predecessor assignment in the ground truth was semantically correct, but syntactically too different to be detectable using our approach.
In 14 cases, we (also) updated the ground truth because we considered the metric-based mapping to be more appropriate.
Afterwards, we applied the same systematic approach to check the 62 posts with false positives or negatives in text blocks.
We noticed that the changes we implemented based on the code block errors also considerably improved the results for text blocks.
For 16 posts, our updated matching strategy removed the false positive and the false negative matches.
Only in 8 text block version comparisons, our strategy was not able to achieve the mapping described in the ground truth, because the predecessor assignment of the text blocks was semantically correct, but syntactically too different to be detectable using our approach.
We updated the ground truth of 41 posts where we considered the metric-based mapping to be more appropriate.
Considering all 83 distinct posts with false positive or false negative matches for either code or text blocks, only eight of them (9.6\%) could not be correctly matched by our revised matching strategy due to the difference between semantic and syntactical difference.
In all other cases, either the revised matching strategy resolved the issues or the ground truth had to be adjusted (dataset available on Zenodo~\cite{BaltesDumaniOthers2017}).
Our next step will be to re-run the complete metrics evaluation (Section~\ref{sec:metrics-evaluation}) to see if, with our revised matching strategy, adjusted thresholds or different metrics yield even better results.

\subsection{Revised Matching Strategy and Post Block Extraction}
\label{sec:revised-strategy}

To address the observed issues, we first changed the \emph{post block extraction} to also detect code blocks that are formatted as inline code, but are the only content in a line and thus formatted as code blocks (see, for example, code block C' in Figure~\ref{fig:matching-strategy-error}).
We further updated the \emph{default similarity metrics} as follows:
We unified the normalization for edit- and n-gram-based metrics and extended the set of special characters by adding colons, commas, and periods.
The reason for this was that $sim_\textit{backup}^\textit{text}$ yielded a similarity of $0.0$ for the strings ``to'' and ``to:'', because they were two different tokens, even after normalization.
We noticed this when checking the false negative matches in question \href{https://stackoverflow.com/posts/38463455/revisions}{38463455} between versions 3 and 4.
In the same post, we further observed a case were the Winnowing algorithm did not trigger the backup metric correctly in case one of the input strings was too short for the configured window size.
We fixed this to resolve the corresponding false negative. 
 
The changes to the \emph{matching strategy} were more complex.
One of the main issues was that we only considered equal predecessors or predecessors with maximum similarity as possible predecessors.
However, those predecessor candidates may not be available anymore at the time our algorithm reaches a certain post block.
Figure~\ref{fig:matching-strategy-error} shows an exemplary false negative match caused by this behavior.
The connection between code blocks \textsc{C'v6} and \textsc{Cv7-1} is missing, because \textsc{Cv7-1} has an equal match in the previous version (\textsc{Cv6}) that is not available anymore at the time the algorithm tries to set the predecessor.
Code block \textsc{C'v6} is very similar to code blocks \textsc{Cv7-1} and \textsc{Cv7-2}, but not equal.
Thus, the set of possible predecessors $Pred(\textsc{Cv7-1})$ only contains \textsc{Cv6}, but not $C'v6$.
We updated the matching strategy as follows to address the above-mentioned issue:

\begin{definition}[Runner-up Predecessors]
Let $b_{(i-1, l)}^{\tau}$, $b_{(i, j)}^{\tau}$ be post blocks of the same type in subsequent post versions and
\[ available(b_{(i-1, l)}^{\tau}) \to \{ \text{true}, \text{false} \} \]
be a function that tests if a post block is still available, meaning that it has not been assigned as predecessor of a post block in the succeeding version yet.
$sim^\tau$, $\vartheta^\tau$, $maxSim^\tau$, and $id^\tau$ have already been defined above.

We define the set of runner-up predecessors as
\begin{align*}
PredRunnerUp(b_{(i, j)}^{\tau})=\{ & \beta^{\tau}_{(i-1, k)}~|~available(\beta^{\tau}_{(i-1, k)}) = true \\
& \land~sim^\tau(\beta^{\tau}_{(i-1, k)} , b_{(i, j)}^{\tau}) \in [ \vartheta^\tau, maxSim^\tau ),\\
& k \in id^\tau(p_{i-1}),~j \in id^\tau(p_{i}) \}
\end{align*}

We define the best runner-up match as
\[ BestRunnerUp(b_{(i, j)}^{\tau}) =
\begin{cases}
\{ b_{(i-1, k)}^{\tau} \} & \mbox{if } \nexists~b_{(i-1, l)}^{\tau} \in PredRunnerUp(b_{(i, j)}^{\tau}): \\
& sim^\tau(b_{(i-1, l)}^{\tau}) > sim^\tau(b_{(i-1, k)}^{\tau}),\\
& k, l \in id^\tau(p_{i-1}),~k \ne l,~j \in id^\tau(p_{i}),\\
\emptyset, & \mbox{else}.
\end{cases} \]
\end{definition}

Using the above definitions, we can now define a new matching strategy that also works in case the optimal match is not available anymore:

$setPredRunnerUp(p_i)$ sets the post block $b_{(i-1, k)}^{\tau} \in BestRunnerUp(b_{(i, j)}^{\tau})$ if $|Succ(b_{(i-1, k)}^{\tau})| = 0$.
Please note that the successor set of $b_{(i-1, k)}^{\tau}$ is empty, because the selected post block did not have the maximum predecessor similarity for any of the post blocks in the succeeding version.
If~$BestRunnerUp(b_{(i, j)}^{\tau}) = \emptyset$, the strategy does not set any predecessors.

Algorithm~\ref{alg:matching-strategy-2} shows the complete revised matching strategy (new parts are marked by a \textit{new} comment).
We use the new matching strategy two times in the algorithm:
At the beginning in case a unique match is not available anymore and in the end after all other strategies were not able to set a predecessor.

\begin{algorithm}
\caption{Revised Matching Strategy}
\label{alg:matching-strategy-2}
\begin{algorithmic}
\ForAll{$p_{2 \le i \le n}$}
	\State \textit{// set predecessors where only one candidate exists}
	\ForAll{$b_{(i, 1 \le j \le |p_i|)}^{\tau}$}
		\If{$|Pred(b_{(i, j)}^{\tau})| = 1$}
			\State Let $pred$ be the equal or similar predecessor
			\If{$available(pred)$} \textit{// new}
				\If{$|Succ(pred)| = 1$}
					\State Set $pred$ as predecessor of $b_{(i, j)}^{\tau}$
					\State \textbf{continue}
				\EndIf
			\Else \textit{// new}
				\State $setPredPositionRunnerUp(p_i)$ \textit{// new}
			\EndIf
		\EndIf
	\EndFor
	\State \textit{// set predecessors using context}
	\State $predSet = \text{true}$
	\While{$predSet$}
		\State $predSet = setPredContext(p_i, BOTH)$
	\EndWhile
	\While{$predSet$}
		\State $predSet = setPredContext(p_i, BELOW)$
	\EndWhile
	\While{$predSet$}
		\State $predSet = setPredContext(p_i, ABOVE)$
	\EndWhile
	\State \textit{// set predecessors using position}
	\State $setPredPosition(p_i)$
	\State \textit{// set runner-up predecessors for the remaining post blocks}
	\State $setPredPositionRunnerUp(p_i)$ \textit{// new}
\EndFor
\end{algorithmic}
\end{algorithm}

\section{Evolution of Stack Overflow Posts}
\label{sec:analysis}


After describing how we reconstructed the version history for individual text and code blocks, we come back to our initial research questions.
We first characterize the phenomenon of SO post evolution, and in particular the evolution of individual post blocks (RQ1).
To find out if edited posts share common characteristics, we analyzed if certain measures such as score or number of comments correlate with the number of edits (RQ2).
We also investigated if those measures have a temporal relationship with the edits, in particular if comments happen immediately before or after edits and whether their relationship follows patterns (RQ3, see Section~\ref{sec:edits-comments}).
Finally, we utilized \emph{SOTorrent} to analyze code clones on SO (RQ4, see Section~\ref{sec:code-clones}).

As descriptive statistics, we use mean ($M$), standard deviation ($SD$), median ($Mdn$), and the first and third quartiles ($Q_1$, $Q_3$).
To test for significant differences, we applied the nonparametric two-sided \textit{Wilcoxon rank-sum test}~\citep{Wilcoxon1945} and report the corresponding p-value ($p_w$).
To measure the effect size, we used \textit{Cohen's} $d$~\citep{Cohen1988, GibbonsHedekerOthers1993}. 
Our interpretation of $d$ is based on the thresholds described by Cohen~\citep{Cohen1992}: negligible effect ($|d|<0.2$), small effect ($0.2\le|d|<0.5$), medium effect ($0.5\le|d|<0.8$), otherwise large effect.
We used the nonparametric \textit{Spearman's rank correlation coefficient} ($\rho$)~\citep{Spearman1904} to test the statistical dependence between two variables.
Our interpretation of $\rho$ is based on Hinkle et al.'s scheme~\citep{HinkleWiersmaOthers1979}: low correlation ($0.3\leq|\rho|<0.5$), moderate correlation ($0.5\leq|\rho|<0.7$), high correlation ($0.7\leq|\rho|<0.9$), and very high correlation ($0.9\leq|\rho|\leq 1$).

\subsection{Quantitative Analysis}
\label{sec:analysis-evolution}

In the following, we describe different properties of post blocks and post block versions either for their most recent version in the \emph{SOTorrent} release 2018-02-16, or for different versions over time:

\paragraph{Post Block Count:}
Half of all posts in the \emph{SOTorrent} dataset contain between one and two text blocks and between zero and two code blocks ($Q_{1,3}$). 
There are only few posts without text blocks ($1.0\%$), but over a third of all posts do not have code blocks ($36.6\%$).
Examples for such posts include conceptual questions and answers, but also posts with inline code that we considered to be part of the text blocks.
If we compare the first and the last version of edited posts, we can observe a statistically significant difference in the number of text and code blocks ($p_w^{\textit{text, code}}<\num{2.2e-16}$); posts tend to grow over time.
However, the effect is only small ($d^\textit{text}=0.21,\; d^\textit{code}=0.23$).

\begin{figure}
\centering
\includegraphics[width=0.88\columnwidth,  trim=0.2in 0.5in 0.2in 0.1in]{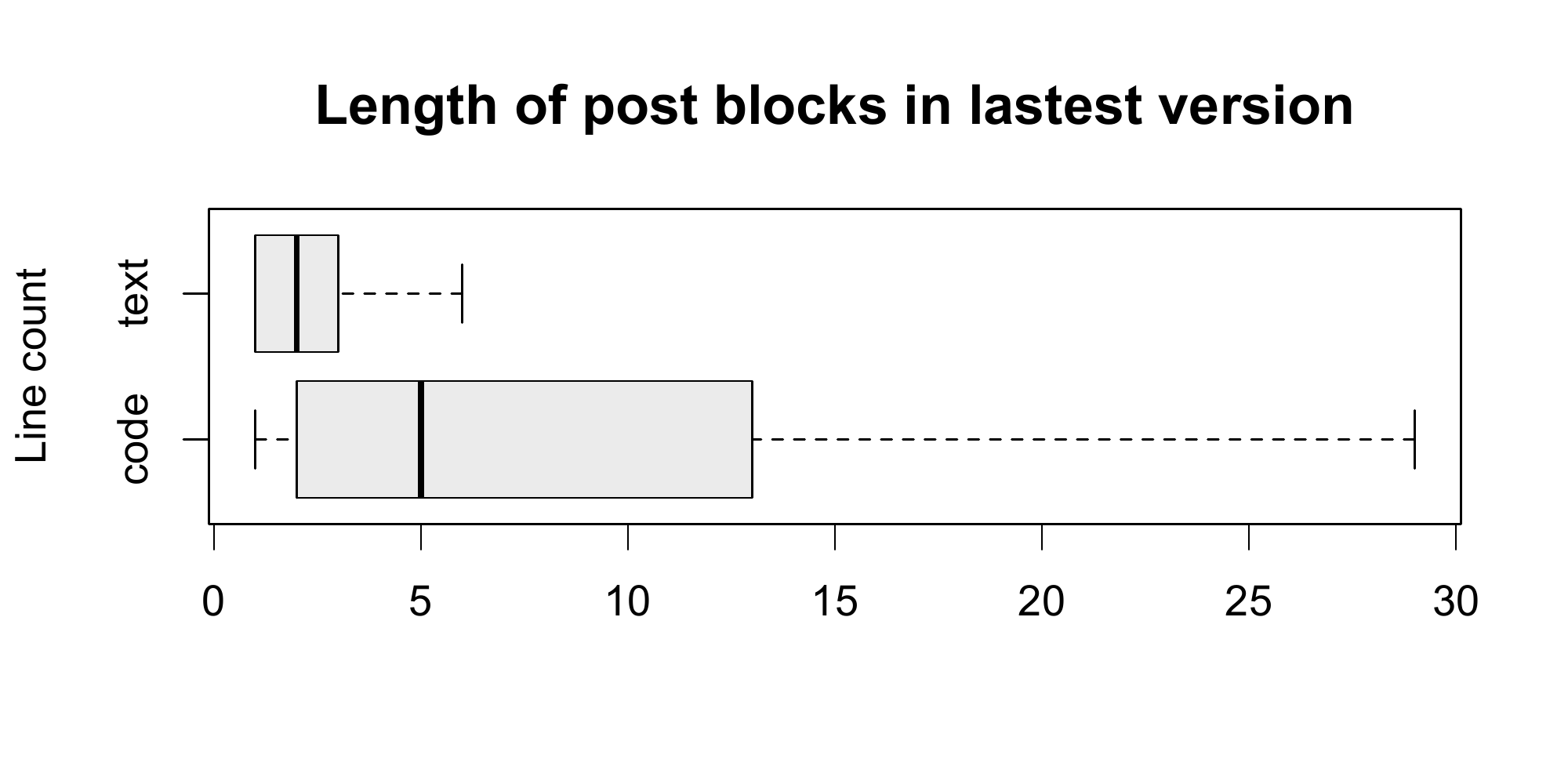} 
\caption{Boxplots showing the line count of text and code blocks in the latest version of Stack Overflow posts ($n=69,940,599$ for text and $n=42,568,011$ for code).}
\label{fig:postblocklength-latest}
\end{figure}

\paragraph{Post Block Length:}
Code blocks tend to be larger than text blocks.
Figure~\ref{fig:postblocklength-latest} visualizes the difference measured in number of lines.
The average text block contains $2.5$ lines ($Mdn=2$, $SD=3.1$) and $247.5$ characters ($Mdn=153$, $SD=319.1$); the average code block contains $12.0$ lines ($Mdn=5$, $SD=23.4$) and $455.9$ characters ($Mdn=194$, $SD=989.3$).
We compared the length of post blocks in the first and the last version and found no effect.
Thus, we can conclude that posts tend to become longer over time in terms of their number of post blocks, but the length of individual post blocks is relatively stable.

\begin{figure}
\centering
\includegraphics[width=0.9\columnwidth,  trim=0.2in 0.6in 0.2in 0.1in]{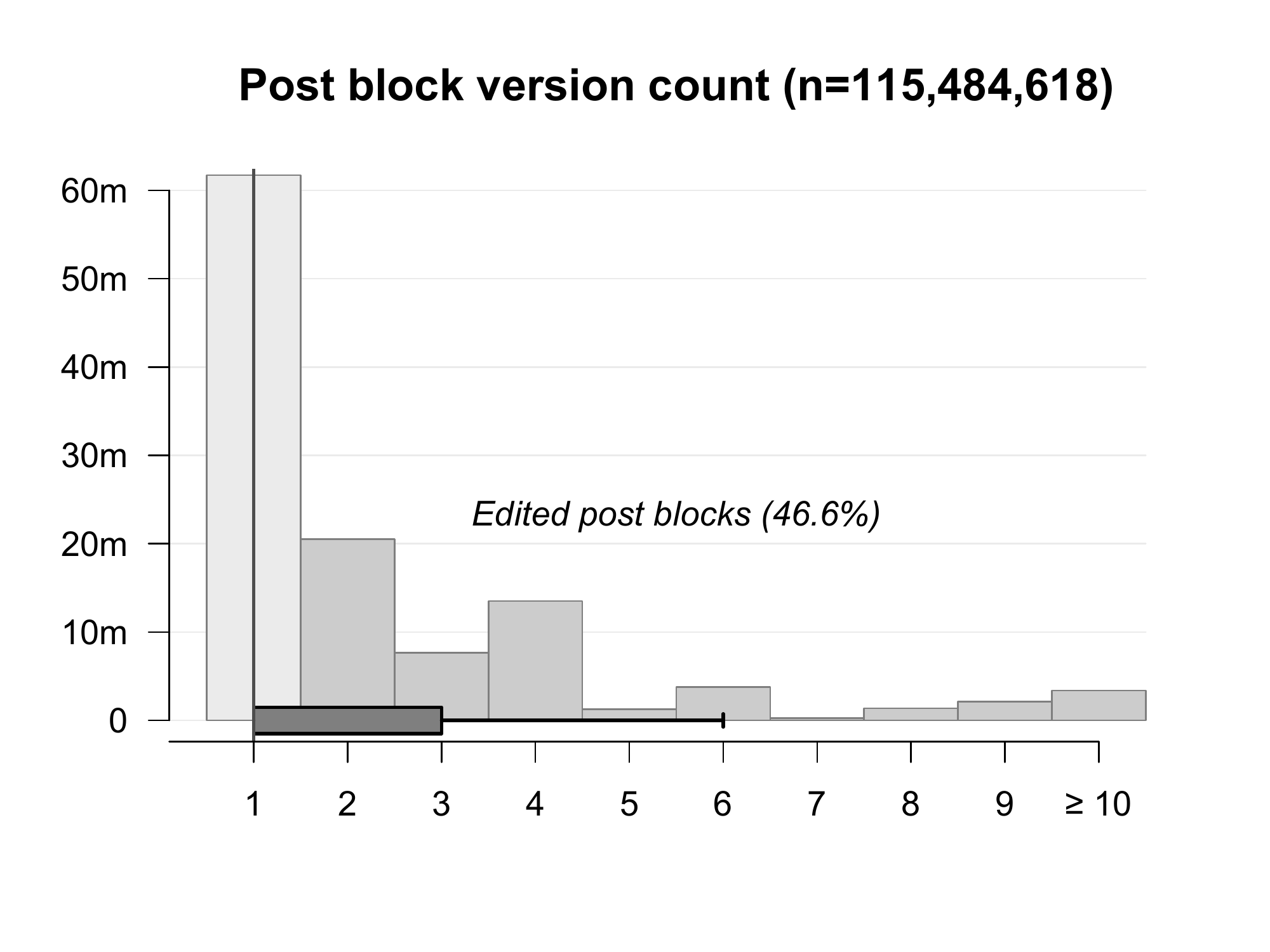} 
\caption{Histogram and boxplot showing the number of post block versions (vertical line visualizes the median value 1).}
\label{fig:postblocklifespan-length}
\end{figure}

\paragraph{Post Block Versions:}
For our analysis of post block versions, we retrieved all post block lifespans in the dataset, but only considered the initial versions and later versions where the content of the blocks changed (not all blocks are edited in all versions).
We found that about half of all post blocks were edited after their creation (see Figure~\ref{fig:postblocklifespan-length}).
On average, text blocks have $4.8$ and code blocks $4.1$ versions.
We analyzed the line-based differences between post block versions and found that $44.1\%$ of all edits modify only one line ($47.7\%$ for text blocks and $34.9\%$ for code blocks).
There is a significant difference in the size of changes when comparing text and code blocks ($p_w<\num{2.2e-16}$) with a medium effect ($d=0.51$ for the number of added lines and $d=0.57$ for the number of deleted lines): 
Changes in code blocks are larger, which is expectable due to the larger size of code compared to text blocks.

\paragraph{Post Block Co-change:}
We were also interested in the co-change of text and code blocks, i.e., if text and code is edited together.
On average, $1.5$ text blocks and $0.9$ code blocks were edited in each post version ($Mdn=1$ and $SD=1.1$ for both types).
We found that text and code blocks were either edited together ($49.3\%$ of all post versions), or just the text blocks were edited ($44.6\%$).
Only in $6.1\%$ of all post versions, code blocks were changed without also editing text blocks.
This could indicate that SO authors document changes to their code snippets in the text blocks or update the description of the modified code.

\begin{figure}
\centering
\includegraphics[width=0.9\columnwidth,  trim=0.0in 0.2in 0.2in 0.1in]{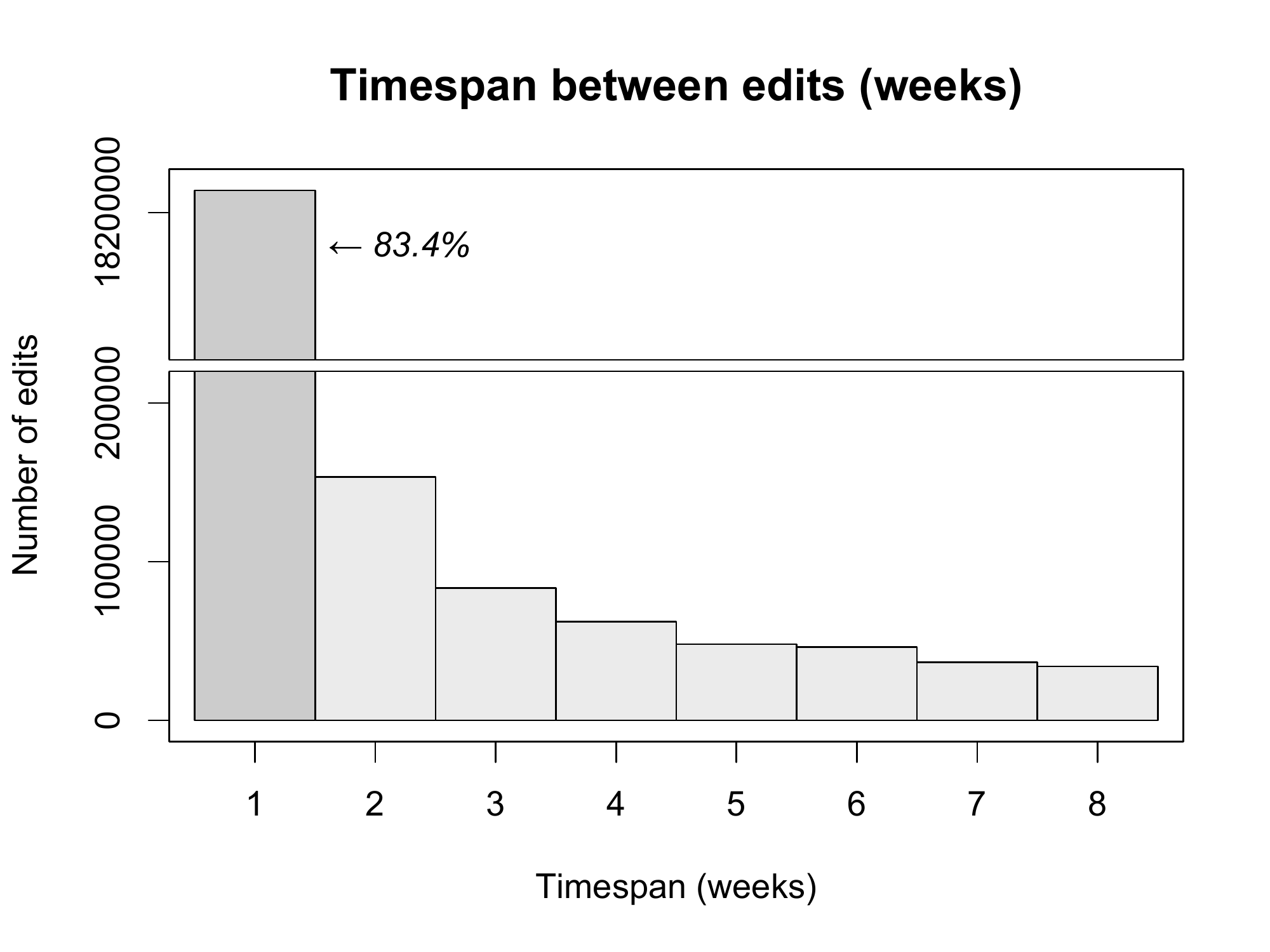} 
\caption{Bar chart visualizing all edit timespans between one and eight weeks ($85.5\%$ of all values, $n=18,677,709$); the other $14.5\%$ are spread over a range of 475 weeks.}
\label{fig:timespan-weeks}
\end{figure}

\paragraph{Order of Post Blocks:}
To check our assumption that the order of post blocks rarely changes, we computed the difference between the local ids of all post blocks versions and their predecessors.
We found that $95.5\%$ of all post block versions have the same local id as their predecessor.
Of all absolute differences, two was the most common one ($3.1\%$), which is expectable, because text and code blocks usually alternate.
Thus, e.g., swapping two blocks of the same type leads to a local id difference of two in the next version.

\paragraph{Timespan Between Edits:}
For the posts that have been edited after their creation, we analyzed the timespan between the edits.
$80.6\%$ of the first post edits happen on the same day as the creation of the post, $4.6\%$ within one week ($>\!1$ and $\le\!7$ days), $5.1\%$ within one year ($>\!7$ and $\le\!365$ days), and $9.7\%$ more than one year after the creation.
If we only consider the second or later edits, not much changes: $74.2\%$ of them happen on the same day, $6.2\%$ within one week, $7.9\%$ within one year, and $11.7\%$ more than one year after the creation.
Overall, $78.2\%$ of all edits happen on the same day, i.e., soon after the creation of the post, and $83.4\%$ happen on the same day or within the first week after the creation (see Figure~\ref{fig:timespan-weeks}).

\paragraph{Post Editors:}
On SO, either the author of a post or a moderator, i.e., a SO user with a reputation of at least 2,000, can make edits.
We found that $87.4\%$ of all edits were conducted by the post authors themselves and $12.6\%$ by moderators.
We found no effect of the authors' reputation on the fact that a moderator edits the post.
We consider an analysis of typical moderator changes to be an interesting direction for future work.

\paragraph{Questions vs. Answers:}
To compare questions and answers, we split the posts according to their post type and then analyzed the three properties \emph{Post Block Count}, \emph{Post Block Length}, and \emph{Post Block Versions} for the most recent version of the posts.
Regarding the post block count, we found that answers have significantly less text and code blocks than questions ($p_w<\num{2.2e-16}$).
The average number of text blocks is $2.1$ ($Mdn=2$ and $SD=1.3$) for questions and $1.6$ ($Mdn=1$ and $SD=1.1$) for answers; the average number of code blocks is $1.3$ ($Mdn=1$ and $SD=1.4$) for questions and $1.0$ ($Mdn=1$ and $SD=1.1$) for answers.
Both effects are small ($d=-0.44$ for text and $d=-0.32$ for code).
We found no effect when comparing the length of text blocks.
However, code blocks in answers tend to be smaller than code blocks in questions.
The average length of answer code blocks was $8.7$ lines ($Mdn=4$ and $SD=16.1$) compared to $15.6$ lines for question code blocks ($Mdn=7$ and $SD=29.0$).
The difference was significant ($p_w<\num{2.2e-16}$) and the effect was small ($d=-0.30$).
The difference is also significant when measured in characters instead of lines ($p_w<\num{2.2e-16}$, $d=-0.31$).
We did not observe a significant difference in the number of versions for questions compared to answers.

\subsection{Properties of Edited Posts}
\label{sec:properties}

\begin{table}
\small
\centering
\caption{Correlation table with Spearman's correlation coefficients $\rho~$ for different properties of Stack Overflow posts (p-value $<0.001$ for all combinations).}
\label{tab:correlations}
\begin{tabular}{c | rrrrr}
\hline
$\rho$ & \multicolumn{1}{c}{Versions} & \multicolumn{1}{c}{Age} & \multicolumn{1}{c}{Score} & \multicolumn{1}{c}{Comments} & \multicolumn{1}{c}{GHMatches} \\
\hline
Versions &  & $-0.03$ & $0.09$ &\graybg\boldsymbol{$0.26$} & $0.09$ \\
Age & $-0.03$ &  & $0.25$ & $-0.03$ & $0.10$ \\
Score & $0.09$ & $0.25$ &  & $0.08$ & $0.23$ \\
Comments &\graybg\boldsymbol{$0.26$} & $-0.03$ & $0.08$ &  & $0.09$ \\
GHMatches & $0.09$ & $0.10$ & $0.23$ & $0.09$ &  \\
\hline
n & 38.4m & 38.4m & 38.4m & 38.4m & 137k \\
\hline
\end{tabular}
\end{table}

To investigate which properties edited posts possess, we searched for monotonic relationships between the version count of a post and other properties such as the age of the post, its score, comment count, or the number of distinct files on GH referring to the post.
Table~\ref{tab:correlations} shows the correlation coefficients ($\rho$) for those relationships based on \emph{SOTorrent} release 2018-02-16.
There was no correlation that exceeded the threshold for a low correlation ($0.3$).
However, the relationship between the version count and the number of comments drew our attention as it had the highest correlation coefficient in the table.
We decided to explore this relationship using a quasi-experiment: We compared the number of comments of all posts with only one version to all posts with more than one version (version count over all posts: $Mdn=1$, $M=1.6$, $SD=1.0$).
The difference was significant ($p_w<\num{2.2e-16}$) and the effect size was medium ($d=0.52$).
We also compared the opposite relationship, i.e., the number of versions of all posts with at most one comment to all posts with more than one comment (comment count over all posts: $Mdn=1$, $M=1.6$, $SD=2.5$).
Again, the difference was significant ($p_w<\num{2.2e-16}$), but the effect size was small ($d=0.49$).

\section{Communication and Edit Patterns}
\label{sec:edits-comments}

Our findings so far suggest that a relationship exists between Stack Overflow post edits and communication events such as comments. To identify common communication and edit patterns in Stack Overflow threads, we first conducted a quantitative analysis of the temporal connection between edits and comments.
A follow-up qualitative study motivated the design of a visual analysis tool that we then used to manually annotate a sample of Stack Overflow threads.

\subsection{Quantitative Analysis}

As a first step in exploring the relationship between comments and post edits, we looked at their temporal connection, i.e., if comments usually happen before or after edits.
First, we aggregated all edits (including post creation) and all comments per post id and day.
Thus, our units of observation were all days where posts were either created, edited or commented.
We found that in $32.3\%$ of the cases, the posts were created or edited and commented; in $33.3\%$ of the cases they were only created, in $9.1\%$ of the cases only edited, in $7.5\%$ of the cases only created and edited, and in $17.8\%$ of the cases only commented.
If we focus on the comments, we see that $64.4\%$ of them happened on a day where the post had either been created or edited.
We then further focused on those days and calculated the time difference between a comment and the closest edit.
If a comment was closer to the creation than to an edit, we assigned the comment to the creation.
We found that $34.7\%$ of the comments were related to the creation of the post and $65.3\%$ were related to an edit.
Of the latter, $47.9\%$ were made before an edit and $52.1\%$ afterwards. 
Moreover, the comments were usually made right before ($M=-1.2~\text{hours}$, $Mdn=-0.3$, $SD=2.6$) or soon after the edits ($M=+1.3~\text{hours}$, $Mdn=+0.3$, $SD=2.7$).

\subsection{Qualitative Analysis}

To further investigate the connection between post edits and comments that are made immediately before or after edits, we conducted a qualitative analysis.
We drew a random sample of 50 posts, 25 posts for which at least one comment had been made at most 10 minutes before an edit and 25 posts for which at least one comment had been made at most 10 minutes after an edit.
We qualitatively analyzed the posts and found that, in the majority of cases, the comments and edits were clearly related (34 of 50 posts in our sample) and that the edit added or modified a code block (30/50).
We classified a small set of comments as bug reports (10/50) and found that in some cases, the edit was explicitly documented in the post (11/50, e.g., by prefixing content with ``\emph{EDIT:}'').
Comments often asked for additional information (22/50), and in cases where comments happened shortly before the edits, the comment was often a clarifying question (14/25).
Answer 15437937\footnote{\url{https://stackoverflow.com/a/15437937}} represents a typical example: In a timespan of 35 minutes, a user answered a question, edited the answer three times, and commented on it once in response to three comments from the user asking the question.
To analyze such communication structures in more detail, we used \emph{SOTorrent} to aggregate edit and comment events for whole threads and built a visual analysis tool to identify patterns.

\subsection{Visual Analysis Tool}

We first aggregated all edit and comment events in \emph{SOTorrent} release 2018-09-23 as described in this blog post.\footnote{\url{http://empirical-software.engineering/blog/sotorrent-edithistory}}
We then drew a random sample of 50 threads with at least one post edit and one comment (see retrieval and analysis scripts on GitHub~\citep{Baltes2018b, Baltes2018c}).
This sample contained 255 edit and 319 comment events from 140 different posts.
We qualitatively analyzed 20 of those threads, which means that we investigated the relationship of 101 edit and 112 comment events from 58 different posts in detail.
To this end, we utilized a web-based visual analysis tool that we specifically designed to analyze the evolution of Stack Overflow threads. Two authors analyzed a subset of the sample and agreed on an annotation strategy, after which one author continued the analysis.

\begin{figure}
\centering
\includegraphics[width=1\columnwidth,  trim=0.0in 0.0in 0.0in 0.0in]{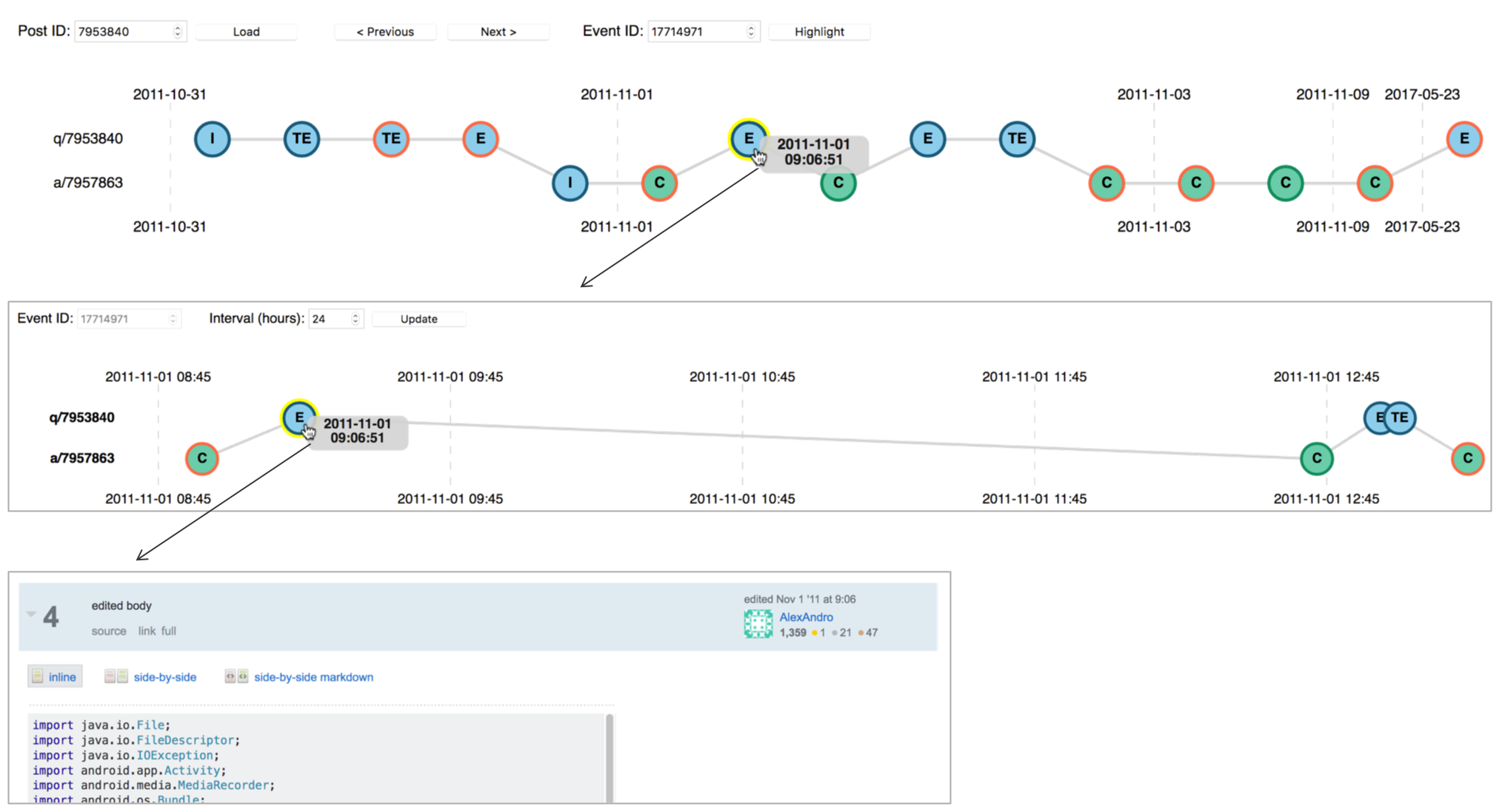} 
\caption{Post evolution visualization: The \texttt{so-edit-viz} \href{http://research.sbaltes.com/so-edit-viz/}{tool} enabled us to visually explore the relationship of edits and comments in Stack Overflow threads (here: thread for question \href{https://stackoverflow.com/q/7953840}{7953840}).}
\label{fig:so-edit-viz-gui}
\end{figure}

Figure~\ref{fig:so-edit-viz-gui} shows the two main views of our visual analysis tool.
The tool provides an overview of the edit and comment events in a thread (upper part of the figure).
It displays the question of a thread in the first row and the answers sorted by their creation date below.
All edit events (I: initial version, E: body edit, TE: title edit) and comment events (C) are plotted using discrete time, with each new day shown as a vertical line.
A circle border in the same color as the circle fill indicates an edit/comment by the post author, a red border indicates an edit/comment by another user.
The currently selected event is highlighted using an additional yellow border, and its event id is also shown in the header.
When hovering over events, a tooltip shows the exact timestamp of the event.
Clicking on an event opens a focused view that uses continuous instead of discrete time, grouped in time frames of one hour (see middle part of Figure~\ref{fig:so-edit-viz-gui} and Figure~\ref{fig:pattern-burst}).
Pressing the `alt' key while clicking on an event in the main view or just clicking on an event in the focused view opens the edit or comment on the Stack Overflow website (see lower part of Figure~\ref{fig:so-edit-viz-gui}).
While the overview enables users to explore the complete evolution of a post, the focused view makes it easier to spot (temporally) related events.
In the example shown in Figure~\ref{fig:so-edit-viz-gui}, the comment and the edit on the left and the agglomeration of edits and comments on the right form two separate groups. 
The source code of the tool together with our remarks for the 20 analyzed threads can be found on GitHub~\citep{Baltes2018g}.
A live demo of the tool is also available.\footnote{\url{http://research.sbaltes.com/so-edit-viz/}}

\subsection{Patterns}

Our analysis revealed six communication and edit patterns, which we describe in the following.

\begin{figure}
\vspace{\baselineskip}
\centering
\includegraphics[width=0.8\columnwidth]{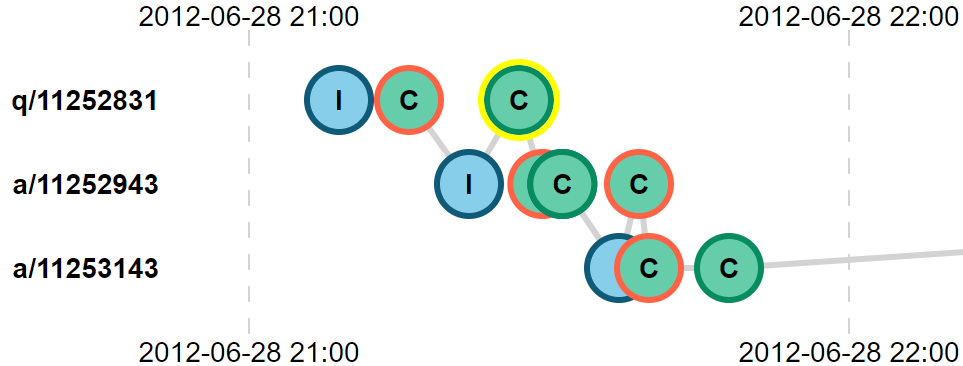} 
\caption{Time line of the burst of commenting and editing activity shortly after Stack Overflow question \href{https://stackoverflow.com/q/11252831}{11252831} was posted.}
\label{fig:pattern-burst}
\end{figure}

\begin{figure}
\vspace{1.5\baselineskip}
\centering
\includegraphics[width=\columnwidth]{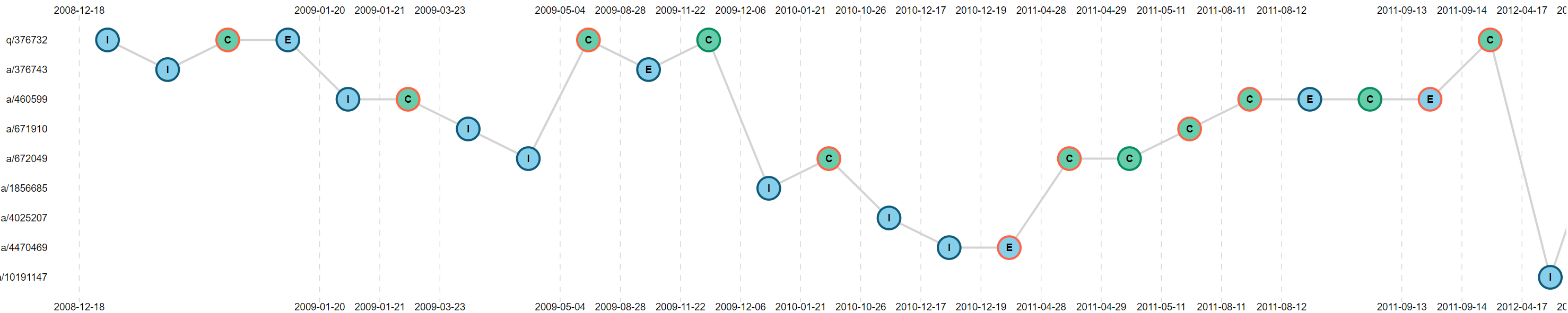} 
\caption{Excerpt of the comment and edit history of Stack Overflow thread \href{https://stackoverflow.com/q/376732}{376732}.}
\label{fig:visualization}
\end{figure}

\paragraph{Burst of Activity:} Several comments and edits occur within minutes of each other. This pattern was very common in our sample of twenty threads: sixteen of the threads contained at least one burst of activity. 

Figure~\ref{fig:pattern-burst} shows part of the time line of Stack Overflow question \href{https://stackoverflow.com/q/11252831}{11252831} to illustrate this pattern: After the initial question was posted, the thread attracted two answers, seven comments, and one post edit within less than 40 minutes. This burst in activity started with a clarification question posted as a comment to the question, followed by the first answer (posted by a different user), and a response to the clarification question by the user who had started the thread. One minute later, the same user then asked a clarification question by commenting on the first answer, in response to which the user who had posted this answer edited it and explained the edit in a comment. Three minutes after that, the third answer was posted, followed by thank-you comments on both answers from the user who had started the thread. Interestingly, the user referred to the edit in the first answer from their comment on the second answer, before the user who had posted the second answer commented that they were planning to update documentation elsewhere to further clarify the issue.

\paragraph{Comment explains Edit:} A comment is used to explain and/or make others aware of an edit. This pattern occurred five times in our sample of twenty threads.

A comment on Stack Overflow question \href{https://stackoverflow.com/q/8687577}{8687577} illustrates this pattern: In response to a jQuery-related question by a new user, another user commented ``1) Welcome to SO. 2) It's not clear what you want to know / are trying to do.'' The user asking the question then proceeded to edit the question to clarify, and left a comment to make the community aware of the edited content: ``I think it should be clearer now [after the] post edit. Thanks again.'' A similar example occurred in Stack Overflow thread \href{https://stackoverflow.com/q/24987992}{24987992}: A user asked a question about how to draw a particular line in D3.js, and another user asked for clarification through a comment: ``Can you post also some image of your wanted output, it's hard to imagine what image you want?'' In response to this comment, the user who had started the thread then edited the question to add a link to an image showing a sketch of the current output and the desired output, and a few minutes later, posted a comment to increase awareness of the edit: ``I upload[ed] the image [url], please take a look''.

\paragraph{Comment triggers Edit:} A post is edited in response to a comment, which happened in four out of the twenty threads in our sample. 

For example, Stack Overflow thread \href{https://stackoverflow.com/q/376732}{376732}, which is visualized in Figure~\ref{fig:visualization}, contains two instances of this pattern: The first comment on the question asks ``What do you have in your .htaccess?'', in response to which the user who had asked the question edited it, adding a six-line code snippet along with the text ``EDIT: This is the current htaccess:''. A similar pattern occurred in the same thread almost three years later: A user commented on the accepted answer, stating ``I don't think it's a valid solution, because with the 404 error you'll be serving the page OK but in the header response you'll see the 404 status code, so it will mess up with your SEO, right?'' The next day, the user who had posted the answer updated it in response to the comment, and also left a new comment explaining the edit (cf.~previous pattern): ``You are right I have changed the example accordingly [...]''.

\paragraph{Question Edit triggers Answer:} An answer is posted shortly after the question has been edited. This pattern occurred twice in our sample.

Stack Overflow thread \href{https://stackoverflow.com/q/13864443}{13864443} is a good example of this pattern. The user who had asked the question did not receive a response right away, and proceeded to make various edits to the question, including the addition of an extra tag and an explanation of the particular constraints of their situation. Within minutes of one of these edits, the first answer to the question was posted -- more than 15 hours after the time the question was originally asked.

\paragraph{Overlap between Comment and Edit:} Text and/or code is copied between comments and post edits, which occurred in two out of the twenty threads in our sample.

In Stack Overflow thread \href{https://stackoverflow.com/q/3529744}{3529744}, the user who had originally asked the question copied a clarification comment they had made in response to another comment into the question text itself: ``It is stand [alone] code. As is. There is no [query] before or after this code.'' A more extreme example of this copy-and-paste pattern occurred in Stack Overflow thread \href{https://stackoverflow.com/q/16245209}{16245209}. The user who had asked the question initially did not include one important code snippet, and was asked for this code snippet in both comments and answers. They then proceeded to edit the question to include a 19-line code snippet, and also added the snippet in form of a comment to the question and the answer. 

\paragraph{Comment announces Edit:} A comment is used to announce a subsequent edit by the same user. We identified two instances of this pattern in our sample.

In both cases, this announcement was made in the context of an ongoing discussion. In Stack Overflow thread \href{https://stackoverflow.com/q/20849332}{20849332}, the user who asked the question commented in responses to a suggestion received in a previous comment on the question: ``[...] I'll update the question in a minute with more detail and some output.'' They proceeded to make the promised edit nine minutes later. In Stack Overflow thread \href{https://stackoverflow.com/q/17591278}{17591278}, the user who had asked the question commented in response to an answer: ``[...] I tried your suggestion with some modification and it worked in a certain way (I'll edit my post in few minutes) [...]'', and the corresponding edit was made less than an hour after this comment.

\section{Code Clones on Stack Overflow}
\label{sec:code-clones}

Code clones have been extensively studied in the software engineering research community.
Juergens et al. found that inconsistent code clones can be a major problem during the development and maintenance of software projects, unless ``special care is taken to find and track existing clones and their evolution''~\citep{JuergensDeissenboeckOthers2009}.
Stack Overflow threads frequently serve as crowd-sourced software documentation~\citep{ParninTreudeOthers2012, TreudeBarzilayOthers2011}, often containing code snippets together with explanations~\citep{YangHussainOthers2016}.
Despite the fact that code clones on Stack Overflow can suffer from similar issues like code clones in software projects, their role has not been investigated yet.
In this section, we present a first analysis of code clones on Stack Overflow, based on the \emph{SOTorrent} dataset.
We will focus on duplicates of code snippets copied from external sources \emph{into SO} and on duplicates of code snippets \emph{within SO}.
The usage and attribution of code snippets copied \emph{from SO} in open source software projects is already covered by our previous work~\citep{BaltesDiehl2018b}.
We were particularly interested in the licensing of snippets copied into Stack Overflow and whether their license status allows redistribution on Stack Overflow.

\subsection{Data Retrieval and Quantitative Analysis}

To detect code clones on Stack Overflow, we utilized the BigQuery version of \emph{SOTorrent} release 2018-09-23.
First, we selected all code blocks from the most recent post versions and normalized the whitespaces.
To this end, we: (1) replaced sequences of new lines with a single new line character, (2) removed new lines at the end of the last line, and (3) removed lines only containing brackets (\texttt{()[]\{\}}).
Using this normalized content, we calculated the normalized line count of those code blocks (NLOC).
Afterwards, we further normalized the content to only contain numbers and digits (character class \texttt{[a-zA-Z0-9]}) and calculated a fingerprint of the normalized code block content using the \texttt{FARM\_FINGERPRINT} function.
This yielded 43,942,960 distinct fingerprints---that is normalized code blocks---in total.
We then used this fingerprint as a \texttt{GROUP BY} argument to determine the posts using a certain snippet, and finally aggregated that information per thread.
The corresponding retrieval script can be found on GitHub~\citep{Baltes2018b}.

As a first filtering step, we selected code blocks that are present in at least two different threads, which was true for 909,323 (2.1\%) of all distinct fingerprints.
Those code clones had an average length of 5.4 normalized lines ($SD=12.9$, $Mdn=2$, $IQR=4$) and were present in 3.5 different threads ($SD=13.3$, $Mdn=2$, $IQR=1$).
To select only non-trivial code snippets, we first used a threshold of six normalized lines of code, as proposed by Bellon et al.~\citep{BellonKoschkeOthers2007}.
We ranked the remaining 215,746 code snippets according to the number of threads they were found in and according to their normalized length.
Then, we qualitatively analyzed the first 50 snippets in that list.
Since we rated 25 of those snippets as non-code (main category were configuration files) or too trivial, we decided to adjust the threshold for the normalized line count to 20.

\begin{figure}
\centering
\includegraphics[width=1\columnwidth,  trim=0.0in 0.4in 0.3in 0.2in]{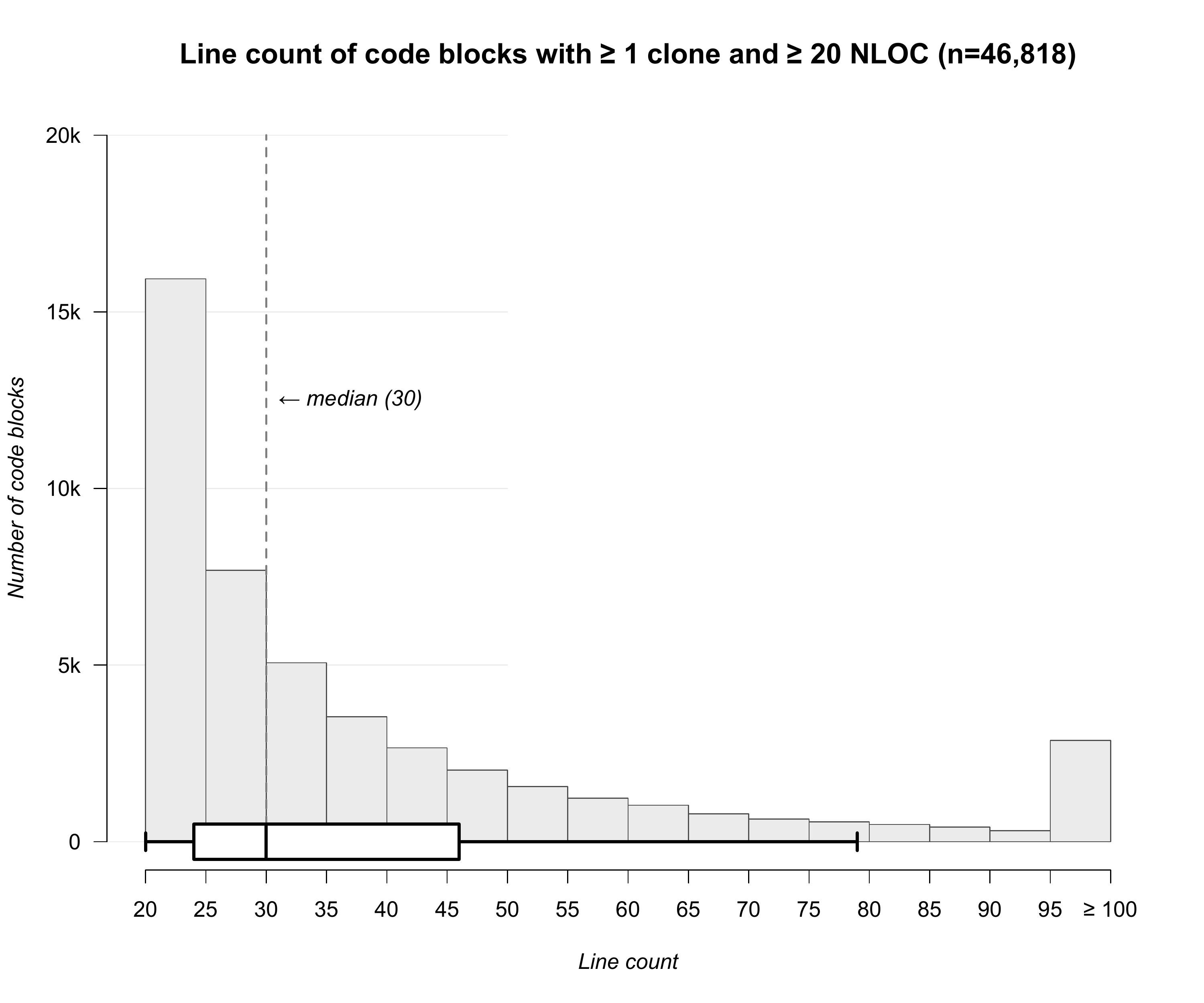} 
\caption{Normalized line count of non-trivial code blocks ($\ge 20$ NLOC) with at least one clone, i.e., present in at least two threads.}
\label{fig:code-clones-line-count}
\end{figure}

\begin{figure}
\centering
\includegraphics[width=1\columnwidth,  trim=0.0in 0.4in 0.3in 0.2in]{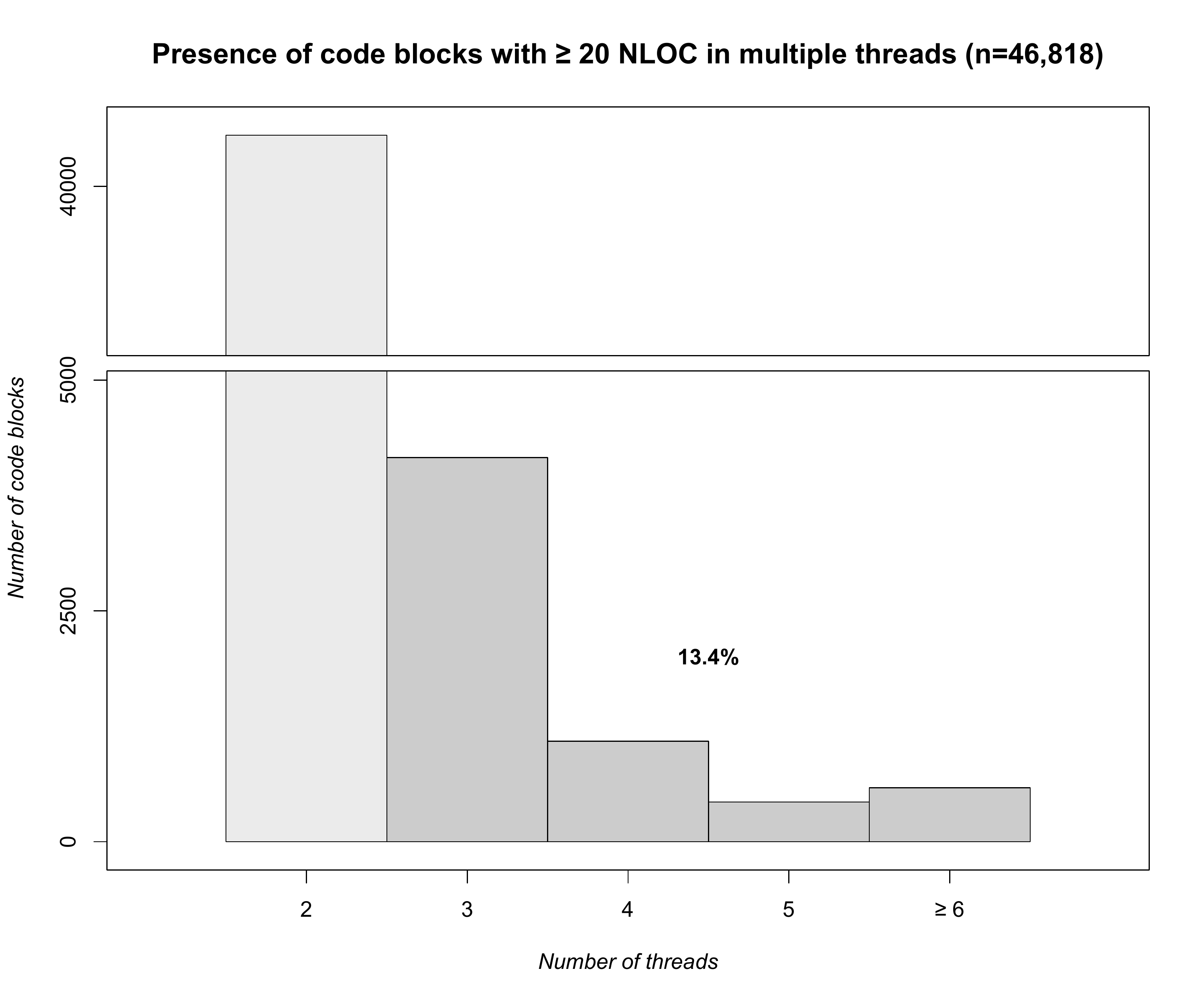} 
\caption{Presence of non-trivial code blocks ($\ge 20$ NLOC) in multiple threads.}
\label{fig:code-clones-thread-count}
\end{figure}

The stricter filtering led to a second sample with 46,818 code snippets.
Those snippets had an average length of 42.6 normalized lines ($SD=37.7$, $Mdn=30$, $IQR=22$) and were present in 2.3 different threads ($SD=1.1$, $Mdn=2$, $IQR=0$)---13.4\% of the snippets were present in more than two threads.
Figures~\ref{fig:code-clones-line-count} and \ref{fig:code-clones-thread-count} visualize the length and thread count distribution in this sample.
We provide the coding for both samples ($\ge 6$ NLOC and $\ge 20$ NLOC) on Zenodo~\citep{Baltes2018f}.
The analysis scripts are available on GitHub~\citep{Baltes2018c}.

\subsection{Qualitative Analysis}

\begin{figure}
\centering
\includegraphics[width=1\columnwidth,  trim=0.0in 0.2in 0.0in 0.0in]{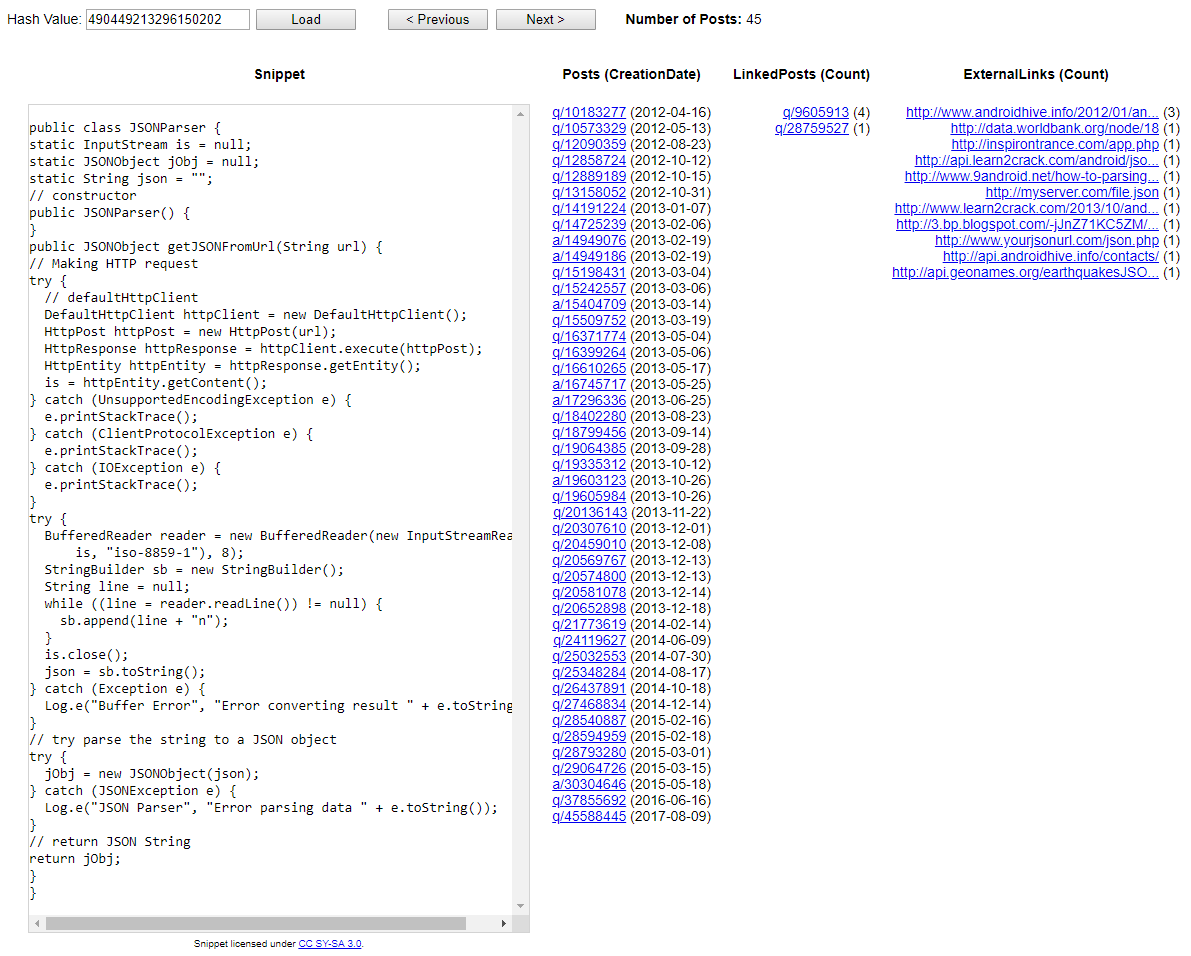} 
\caption{Snippet view of \texttt{so-clones} \href{http://research.sbaltes.com/so-clones/snippet-view.html?hashValue=490449213296150202}{tool} showing a code snippet that has likely been copied from the website \href{https://www.androidhive.info/2012/05/how-to-connect-android-with-php-mysql/}{\emph{androidhive}} into Stack Overflow.}
\label{fig:so-clones-snippet-view}
\end{figure}

In the second sample, we again ranked the code snippets according to their thread count and length to qualitatively analyze the first 50 snippets in the resulting list.
We also implemented a web tool\footnote{\url{http://research.sbaltes.com/so-clones/}} that allowed us to explore that list.
The tool enables users not only to browse the complete list, but it also to focus on  a single snippet in a dedicated view.
This view (see Figure~\ref{fig:so-clones-snippet-view}) shows the snippet, its fingerprint, the posts containing the snippet sorted by their creation date, other Stack Overflow posts linked from those posts, and linked external sources.
The latter information helped us a lot in identifying if and from where a snippet has been copied into Stack Overflow.
The source code of the tool is available on GitHub~\citep{Baltes2018f}.
 
While there were still ten snippets that we categorized as configuration files, 29 snippets were non-trivial source code snippets (mainly Java and VB/VBA).
Other categories included XML GUI definitions for Android, JSON/XML examples, and HTML files.
Except for two cases, we were able to identify the (or at least \emph{a}) source of the snippet.
Only in four cases, we considered the snippets to be originally from Stack Overflow.
The main external sources were a website providing Android tutorials\footnote{\url{https://www.androidhive.info/}} (ten snippets) and the official Android documentation\footnote{\url{http://developer.android.com/}} (4 snippets).
We identified a possible licensing conflict in 31 cases, either because the website did not provide any license or because the content was distributed under a restrictive license or under restrictive terms of use.
In the following, we are going to describe the two main external sources in more detail.
The independent Android website \emph{androidhive} has rather restrictive terms of use\footnote{\url{https://www.androidhive.info/terms-of-service/}}:

\begin{quote}
\emph{``Our Website is also protected under international copyright laws.
The copying, redistribution, use or publication by you of any portion of our Website is strictly prohibited.''}
\end{quote}

Nevertheless, only few Stack Overflow posts attribute this source (3 out of 45 posts in the example shown in Figure~\ref{fig:so-clones-snippet-view}).
It is unclear if the snippet has actually been copied from this external source (the creation of the posts on \emph{androidhive} and Stack Overflow was around April/May 2012).
But if this is the case, the 45 snippets copied on Stack Overflow could be problematic.
In fact, we identified four more variants of that same code snippet among the 50 snippets we analyzed.
On the other hand, if Stack Overflow is the original source, the usage on \emph{androidhive} does not adhere to Stack Overflow's CC BY-SA license~\citep{BaltesDiehl2018b}.

The snippets copied from the official Android documentation are licensed under CC BY 2.5\footnote{\url{https://developer.android.com/license}}.
This license allows usage under Stack Overflow's  CC BY-SA license, but only when attributing the original source.
However, only in few cases the users added a link to the Android documentation to their posts.
Thus, also those usages could lead to \emph{licensing issues}.

Leaving the licensing implications aside, the code clones within Stack Overflow are also problematic for the platform's \emph{usability}.
Those duplicates could indicate that different threads solved a similar problem.
However, if there is no link between those threads, the information is spread and hard to capture for readers.
Stack Overflow recommends to \emph{``always quote the most relevant part of an important link, in case the target site is unreachable or goes permanently offline''}.\footnote{\url{https://stackoverflow.com/help/how-to-answer}}
While it makes sense to quote the main points of an external source or pseudo code of algorithms, it can be questioned if it is reasonable to have several independent copies of non-trivial code snippets on Stack Overflow.
Assuming the snippet in the reference documentation is updated, all copies on Stack Overflow (14 in \href{http://research.sbaltes.com/so-clones/snippet-view.html?hashValue=-4458403169525607496}{this} example) must also be updated.
Again, only few Stack Overflow authors link to other posts that already provided the same snippet, making it even harder to update them.

To discuss how to best approach those licensing and usability issues, we created a post on Stack Overflow Meta~\citep{StackOverflowMeta2018} to involve the community.
We asked, for example, if it would make sense to point Stack Overflow users to related threads based on the similarity of the code blocks posted in a thread, which could be done before users post a question or integrated into the website for existing posts.
The post got upvoted to a score of 28 (as of October 30, 2018) and is being discussed in the comments, but there is no answer yet.
Stack Overflow user Martijn Pieters, for example, wrote\footnote{\url{https://meta.stackoverflow.com/questions/375761/how-to-handle-code-clones-on-stack-overflow\#comment641119_375761}}:

\begin{quote}
\emph{``I see this a lot in Java (especially Android) code when researching serial plagiarists.
There is a lot of example code floating around that is free to copy, but there seems to be an endemic culture that sees copying as a legitimate method of developing software.
[...]
answers should primarily be your own work, not someone else's.''}
\end{quote}

One preliminary result of the discussion is that there are comments in favor of adding the missing attribution for the external source to the Stack Overflow posts.
However, this would only solve the licensing issue for snippets licensed under a rather permissive license.
Moreover, the clones on Stack Overflow would still be isolated from each other.
Depending on the outcome of the discussion on Stack Overflow Meta, we plan to implement the approach that the community favors, for example by automatically proposing post edits to add the missing attribution.

\section{Discussion}

The \emph{SOTorrent} dataset has allowed us to study the phenomenon of post editing on SO in detail (RQ1).
We found that a total of 13.9 million SO posts ($36.1\%$ of all posts) have been edited at least once.
Many of these edits ($44.1\%$) modify only a single line of text or code, and while posts grow over time in terms of the number of text and code blocks they contain, the size of these individual blocks is relatively stable.
Interestingly, only in $6.1\%$ of all cases are code blocks changed without corresponding changes in text blocks of the same post, suggesting that SO users typically update the textual description accompanying code snippets when they are edited.
We also found that edits are mostly made shortly after the creation of a post ($78.2\%$ of all edits are made on the same day when the post was created), and the vast majority of edits are made by post authors ($87.4\%$)---although the remaining $12.6\%$ will be of particular interest for our future work.
The number of comments on posts without edits is significantly smaller than the number of comments on posts with edits, suggesting an interplay of these two features (RQ2).
We find evidence which suggests that commenting on a post on SO helps to bring attention to it (RQ3).
Of the comments that were made on the same day as an edit, $47.9\%$ were made before an edit and $52.1\%$ afterwards, typically (median value) only 18 minutes before or after the edit.

Motivated by this quantitative analysis of the temporal relationship between edits and comments, we conducted a qualitative study and developed a visual analysis tool to explore the communication structure of Stack Overflow threads.
Our analysis using this tool revealed several communication and edit patterns (RQ3) that provide further evidence for the connection between post edits and comments.
We found comments which explain, trigger, and announce edits as well as content overlap between edits and comments.
The fact that Stack Overflow users rely on the commenting feature to make others aware of post edits---and in some cases even duplicate content between comments and posts---suggests that users are worried that content evolution will be missed if it is buried in a comment or has been added to a post later via an edit.
At the same time, we found evidence that edits can play a vital role in attracting answers to a question.
In our future work, we will explore how changes to Stack Overflow's user interface could make the evolution of content more explicit and remove the need for users to repurpose the commenting feature as an awareness mechanism.

Besides, we presented a first investigation of code clones on Stack Overflow (RQ4) that revealed that, just like in regular software projects, code clones on Stack Overflow can affect the maintainability of posts and lead to licensing issues.
Depending on the outcome of the discussion we started on Stack Overflow Meta, we plan to implement means to add the missing attribution to posts and mark threads as related based on the similarity of the code blocks they contain.

\section{Related Work}

Over the past years, there have been various research papers on leveraging knowledge from SO, e.g., to support post edits~\citep{ChenXingOthers2017}, to automate the search~\citep{PonzanelliBacchelliOthers2013, CampbellTreude2017}, or to augment API documentation~\citep{TreudeRobillard2016}.
Regarding the population of SO users, studies described properties such as gender~\citep{VasilescuCapiluppiOthers2012} and age~\citep{MorrisonMurphyHill2013}.
Wang et al.~\citep{WangLoDavidOthers2013} analyzed the asking and answering behavior of SO users and found that most of them only answer or ask one question.
We complement those results with our finding that post edits happen soon after post creation and that comments are closely linked to edits.
Xia et al.~\citep{XiaBaoOthers2017} describe that it is common for developers to search for reusable code snippets on the web.
Yang et al.~\citep{YangHussainOthers2016} found that SO Python and JavaScript snippets are more usable in terms of parsability, compilability and runnability, compared to Java and C\#.
Yang et al.~\citep{YangMartinsOthers2017} analyzed code clones between Python snippets from SO and Python projects on GH and found a considerable number of non-trivial clones, which may have a negative impact on code quality~\citep{AbdalkareemShihabOthers2017}.
Baltes and Diehl~\citep{BaltesDiehl2018b} investigated the usage and attribution of SO code snippets in GH projects and found that at most a quarter of the usages are attributed as required by SO's license.
Moreover, they point to possible licensing issues, similar to what we described in Section~\ref{sec:code-clones}.
Other studies aimed at identifying API usage in SO code snippets~\citep{SubramanianHolmes2013}, describing characteristics of effective code examples~\citep{NasehiSillitoOthers2012}, investigating whether SO code snippets are self-explanatory~\citep{TreudeRobillard2017}, or analyzing the impact of copied SO code snippets on application security~\citep{AcarBackesOthers2016, FischerBottingerOthers2017}.
There has also been work on the interplay between user activity on SO and GH~\citep{VasilescuFilkovOthers2013, SilvestriYangOthers2015, BadashianEstekiOthers2014}.
\emph{SOTorrent} enables researchers to further investigate this connection by collecting links from public GH projects to SO posts. 
To describe topics of SO questions and answers, different methods like manual analysis~\citep{TreudeBarzilayOthers2011} and Latent Dirichlet Allocation~\citep{WangLoDavidOthers2013, AllamanisSutton2013} have been used.
Automatically identifying high-quality posts has been another research direction, where metrics based on the number of edits on a question~\citep{YangHauffOthers2014}, author popularity~\citep{PonzanelliMocciOthers2014}, and code readability~\citep{DuijnKuceraOthers2015} yielded good results.
With our dataset, the evolution of such high-quality posts can easily be analyzed.
German et al.~\citep{GermanDiPentaOthers2009} investigated how code siblings, code clones that evolve in a different system than the original code, flow between systems with different licenses.
Tracing the flow of siblings between GH projects, posts on SO, and external sources is another possible direction for future work that \emph{SOTorrent} can support.
Two fields related to our research are source code plagiarism detection~\citep{LancasterCulwin2004} and code clone detection~\citep{RoyCordyOthers2009}, which both rely on determining the similarity of code fragments.

\section{Conclusion}

In this paper, we described how we reconstructed and analyzed the evolution of Stack Overflow posts.
We presented the open dataset \emph{SOTorrent} that enables researchers to analyze the evolution of SO content at the level of whole posts and individual text and code blocks.
We described how we evaluated 134 different string similarity metrics regarding their suitability to match text and code blocks to their predecessor versions.
For text blocks, a profile-based metric using the Manhattan distance yielded the best results; for code blocks, a fingerprint-based metric using the Winnowing algorithm~\citep{SchleimerWilkersonOthers2003, DuricGasevic2013} outperformed the other metrics.
Since multiple predecessor candidates may exist, we also developed a matching strategy that we iteratively refined using random samples of SO posts.
After an analysis of false positive and false negative matches, we further improved this strategy.

Our quantitative and qualitative analyses based on the dataset provided new insights into the evolution of SO posts, and in particular the relationship between post edits and comments and the presence of code clones on SO.
In future work, we want to deepen our understanding of how code snippets are maintained on SO, and how code clones affect their maintainability.
Moreover, as \emph{SOTorrent} also collects links from SO posts to other websites and from public GH projects to SO posts, we want to explore how code flows from and to external sources like blog posts and open source software projects.
Beside the investigation of new research questions, we will continue to improve and maintain the dataset, for example by developing means to automatically detect code blocks that are not used for code, but for markup (see, e.g., second code block in Figure~\ref{fig:so-postblocks-example}).
We welcome bug reports and ideas for improvements, especially by researchers who use \emph{SOTorrent} to investigate the evolution of SO posts and their connection to other platforms and resources.
Everyone can provide feedback simply by creating an issue on GitHub.\footnote{\url{https://github.com/sotorrent/db-scripts/issues}}



\begin{acknowledgements}
The authors would like to thank Florian Reitz for his help with database-related issues and Tobias Zeimetz for creating the post history ground truth.
\end{acknowledgements}

\bibliographystyle{spbasic}      
\bibliography{literature}   

\begin{thebibliography}{73}
\providecommand{\natexlab}[1]{#1}
\providecommand{\url}[1]{{#1}}
\providecommand{\urlprefix}{URL }
\expandafter\ifx\csname urlstyle\endcsname\relax
  \providecommand{\doi}[1]{DOI~\discretionary{}{}{}#1}\else
  \providecommand{\doi}{DOI~\discretionary{}{}{}\begingroup
  \urlstyle{rm}\Url}\fi
\providecommand{\eprint}[2][]{\url{#2}}

\bibitem[{Abdalkareem et~al.(2017)Abdalkareem, Shihab, and
  Rilling}]{AbdalkareemShihabOthers2017}
Abdalkareem R, Shihab E, Rilling J (2017) {On code reuse from StackOverflow: An
  exploratory study on Android apps}. {Information and Software Technology}
  88:148--158

\bibitem[{Acar et~al.(2016)Acar, Backes, Fahl, Kim, Mazurek, and
  Stransky}]{AcarBackesOthers2016}
Acar Y, Backes M, Fahl S, Kim D, Mazurek ML, Stransky C (2016) {You Get Where
  You're Looking For: The Impact Of Information Sources on Code Security}. In:
  Locasto M, Shmatikov V, Erlingsson {\'U} (eds) {2016 IEEE Symposium on
  Security and Privacy (S{\&}P 2016)}, {IEEE Computer Society}, San Jose, CA,
  USA, pp 289--305

\bibitem[{Allamanis and Sutton(2013)}]{AllamanisSutton2013}
Allamanis M, Sutton C (2013) {Why, when, and what: Analyzing Stack Overflow
  questions by topic, type, and code}. In: Zimmermann T, {Di Penta} M, Kim S
  (eds) {10th International Working Conference on Mining Software Repositories
  (MSR 2013)}, IEEE, San Francisco, CA, USA, pp 53--56

\bibitem[{An et~al.(2017)An, Mlouki, Khomh, and Antoniol}]{AnMloukiOthers2017}
An L, Mlouki O, Khomh F, Antoniol G (2017) {Stack Overflow: A Code Laundering
  Platform?} In: Pinzger M, Bavota G, Marcus A (eds) {24th IEEE International
  Conference on Software Analysis, Evolution and Reengineering (SANER 2017)},
  {IEEE Computer Society}, Klagenfurt, Austria, pp 283--293

\bibitem[{Badashian et~al.(2014)Badashian, Esteki, Gholipour, Hindle, and
  Stroulia}]{BadashianEstekiOthers2014}
Badashian AS, Esteki A, Gholipour A, Hindle A, Stroulia E (2014) {Involvement,
  Contribution and Influence in GitHub and Stack Overflow}. In: Ng J, Li J,
  Wong K (eds) {24th International Conference on Computer Science and Software
  Engineering (CASCON 2014)}, {IBM / ACM}, Markham, ON, Canada, pp 19--33

\bibitem[{Baltes(2018{\natexlab{a}})}]{Baltes2018d}
Baltes S (2018{\natexlab{a}}) {SOTorrent: Reconstructing and Analyzing the
  Evolution of Stack Overflow Posts --- Supplementary Material}.
  \urlprefix\url{http://doi.org/10.5281/zenodo.1201553}

\bibitem[{Baltes(2018{\natexlab{b}})}]{Baltes2018b}
Baltes S (2018{\natexlab{b}}) {sotorrent/db-scripts on GitHub}.
  \urlprefix\url{https://doi.org/10.5281/zenodo.1116346}

\bibitem[{Baltes(2018{\natexlab{c}})}]{Baltes2018c}
Baltes S (2018{\natexlab{c}}) {sotorrent/r-scripts on GitHub}.
  \urlprefix\url{https://doi.org/10.5281/zenodo.1048185}

\bibitem[{Baltes(2018{\natexlab{d}})}]{Baltes2018f}
Baltes S (2018{\natexlab{d}}) {sotorrent/so-clones on GitHub}.
  \urlprefix\url{https://doi.org/10.5281/zenodo.1472948}

\bibitem[{Baltes(2018{\natexlab{e}})}]{Baltes2018g}
Baltes S (2018{\natexlab{e}}) {sotorrent/so-edit-viz on GitHub}.
  \urlprefix\url{https://doi.org/10.5281/zenodo.1474203}

\bibitem[{Baltes(2018{\natexlab{f}})}]{Baltes2018e}
Baltes S (2018{\natexlab{f}}) {Usage and Attribution of Stack Overflow Code
  Snippets in GitHub Projects --- Supplementary Material}.
  \urlprefix\url{https://doi.org/10.5281/zenodo.1148069}

\bibitem[{Baltes and Diehl(2018)}]{BaltesDiehl2018b}
Baltes S, Diehl S (2018) {Usage and Attribtion of Stack Overflow Code Snippets
  in GitHub Projects}. {Empirical Software Engineering} Online First:1--37

\bibitem[{Baltes and Dumani(2018{\natexlab{a}})}]{BaltesDumani2018}
Baltes S, Dumani L (2018{\natexlab{a}}) {SOTorrent Data Set Version
  2018-02-16}. \urlprefix\url{http://doi.org/10.5281/zenodo.1196296}

\bibitem[{Baltes and Dumani(2018{\natexlab{b}})}]{BaltesDumani2018b}
Baltes S, Dumani L (2018{\natexlab{b}}) {SOTorrent GitHub Page}.
  \urlprefix\url{https://github.com/sotorrent}

\bibitem[{Baltes and Dumani(2018{\natexlab{c}})}]{BaltesDumani2018c}
Baltes S, Dumani L (2018{\natexlab{c}}) {sotorrent/metrics-comparison on
  GitHub}. \urlprefix\url{https://doi.org/10.5281/zenodo.1045823}

\bibitem[{Baltes and Dumani(2018{\natexlab{d}})}]{BaltesDumani2018d}
Baltes S, Dumani L (2018{\natexlab{d}}) {sotorrent/so-posthistory-extractor on
  GitHub}. \urlprefix\url{https://doi.org/10.5281/zenodo.835046}

\bibitem[{Baltes et~al.(2017{\natexlab{a}})Baltes, Dumani, and
  Zeimetz}]{BaltesDumaniOthers2017}
Baltes S, Dumani L, Zeimetz T (2017{\natexlab{a}}) {Dataset with manually
  validated version histories of Stack Overflow posts}.
  \urlprefix\url{http://doi.org/10.5281/zenodo.884909}

\bibitem[{Baltes et~al.(2017{\natexlab{b}})Baltes, Kiefer, and
  Diehl}]{BaltesKieferOthers2017}
Baltes S, Kiefer R, Diehl S (2017{\natexlab{b}}) {Attribution required: Stack
  overflow code snippets in GitHub projects}. In: Uchitel S, Orso A, Robillard
  MP (eds) {39th International Conference on Software Engineering (ICSE 2017),
  Companion Volume}, {IEEE Computer Society}, Buenos Aires, Argentina, pp
  161--163

\bibitem[{Baltes et~al.(2018)Baltes, Dumani, Treude, and
  Diehl}]{BaltesDumaniOthers2018}
Baltes S, Dumani L, Treude C, Diehl S (2018) {SOTorrent: Reconstructing and
  Analyzing the Evolution Stack Overflow Posts}. In: Zaidman A, Hill E, Kamei Y
  (eds) {15th International Conference on Mining Software Repositories (MSR
  2018)}, ACM, Gothenburg, Sweden, pp 319--330

\bibitem[{Bellon et~al.(2007)Bellon, Koschke, Antoniol, Krinke, and
  Merlo}]{BellonKoschkeOthers2007}
Bellon S, Koschke R, Antoniol G, Krinke J, Merlo E (2007) {Comparison and
  evaluation of clone detection tools}. {IEEE Transactions on Software
  Engineering} 33(9):577--591

\bibitem[{Burrows et~al.(2007)Burrows, Tahaghoghi, and
  Zobel}]{BurrowsTahaghoghiOthers2007}
Burrows S, Tahaghoghi SMM, Zobel J (2007) {Efficient plagiarism detection for
  large code repositories}. {Software---Practice and Experience} 37(2):151--176

\bibitem[{Campbell and Treude(2017)}]{CampbellTreude2017}
Campbell BA, Treude C (2017) {NLP2Code: Code Snippet Content Assist via Natural
  Language Tasks}. In: Mei H, Zhang L, Zimmermann T (eds) {2017 IEEE
  International Conference on Software Maintenance and Evolution (ICSME 2017)},
  {IEEE Computer Society}, Shanghai, China, pp 628--632

\bibitem[{Chapin et~al.(2001)Chapin, Hale, Khan, Ramil, and
  Tan}]{ChapinHaleOthers2001}
Chapin N, Hale JE, Khan KM, Ramil JF, Tan WG (2001) {Types of software
  evolution and software maintenance}. {Journal of Software Maintenance}
  13(1):3--30

\bibitem[{Chen et~al.(2017)Chen, Xing, and Liu}]{ChenXingOthers2017}
Chen C, Xing Z, Liu Y (2017) {By the Community {\&} For the Community: A Deep
  Learning Approach to Assist Collaborative Editing in Q{\&}A Sites}.
  {Proceedings of the ACM on Human-Computer Interaction} 1:32:1--32:21

\bibitem[{Chicco(2017)}]{Chicco2017}
Chicco D (2017) {Ten quick tips for machine learning in computational biology}.
  {BioData mining} 10(1):35

\bibitem[{Cohen(1988)}]{Cohen1988}
Cohen J (1988) {Statistical Power Analysis for the Behavioral Sciences}, 2nd
  edn. Routledge, Mahwah, NJ, USA

\bibitem[{Cohen(1992)}]{Cohen1992}
Cohen J (1992) {A power primer}. {Psychological bulletin} 112(1):155

\bibitem[{Duijn et~al.(2015)Duijn, Kucera, and
  Bacchelli}]{DuijnKuceraOthers2015}
Duijn M, Kucera A, Bacchelli A (2015) {Quality Questions Need Quality Code:
  Classifying Code Fragments on Stack Overflow}. In: {Di Penta} M, Pinzger M,
  Robbes R (eds) {12th Working Conference on Mining Software Repositories (MSR
  2015)}, {IEEE Computer Society}, Florence, Italy, pp 410--413

\bibitem[{Dumani and Baltes(2017)}]{DumaniBaltes2017}
Dumani L, Baltes S (2017) {sotorrent/so-posthistory-gt on GitHub}.
  \urlprefix\url{https://doi.org/10.5281/zenodo.1045935}

\bibitem[{Dumani and Baltes(2018)}]{DumaniBaltes2018}
Dumani L, Baltes S (2018) {sotorrent/posthistory-comparator-gt-cs on GitHub}.
  \urlprefix\url{https://doi.org/10.5281/zenodo.1474238}

\bibitem[{Duric and Gasevic(2013)}]{DuricGasevic2013}
Duric Z, Gasevic D (2013) {A source code similarity system for plagiarism
  detection}. {The Computer Journal} 56(1):70--86

\bibitem[{Fischer et~al.(2017)Fischer, B{\"o}ttinger, Xiao, Stransky, Acar,
  Backes, and Fahl}]{FischerBottingerOthers2017}
Fischer F, B{\"o}ttinger K, Xiao H, Stransky C, Acar Y, Backes M, Fahl S (2017)
  {Stack Overflow Considered Harmful? The Impact of Copy{\&}Paste on Android
  Application Security}. In: Butler KRB, Erlingsson {\'U}, Parno B (eds) {2017
  IEEE Symposium on Security and Privacy (S{\&}P 2017)}, {IEEE Computer
  Society}, San Jose, CA, USA, pp 121--136

\bibitem[{German et~al.(2009)German, {Di Penta}, Gueheneuc, and
  Antoniol}]{GermanDiPentaOthers2009}
German DM, {Di Penta} M, Gueheneuc YG, Antoniol G (2009) {Code siblings:
  Technical and legal implications of copying code between applications}. In:
  Godfrey MW, Whitehead J (eds) {6th International Working Conference on Mining
  Software Repositories (MSR 2009)}, {IEEE Computer Society}, Vancouver, BC,
  Canada, pp 81--90

\bibitem[{Gharehyazie et~al.(2017)Gharehyazie, Ray, and
  Filkov}]{GharehyazieRayOthers2017}
Gharehyazie M, Ray B, Filkov V (2017) {Some From Here, Some From There:
  Cross-Project Code Reuse in GitHub}. In: Gonzalez-Barahona JM, Hindle A, Tan
  L (eds) {14th International Conference on Mining Software Repositories (MSR
  2017)}, {IEEE Computer Society}, Buenos Aires, Argentina, pp 291--301

\bibitem[{Gibbons et~al.(1993)Gibbons, Hedeker, and
  Davis}]{GibbonsHedekerOthers1993}
Gibbons RD, Hedeker DR, Davis JM (1993) {Estimation of effect size from a
  series of experiments involving paired comparisons}. {Journal of Educational
  Statistics} 18(3):271--279

\bibitem[{Godfrey and German(2008)}]{GodfreyGerman2008}
Godfrey MW, German DM (2008) {The past, present, and future of software
  evolution}. In: Muller H, Tilley S, Wong K (eds) {Frontiers of Software
  Maintenance (FoSM 2008)}, IEEE, Beijing, China, pp 129--138

\bibitem[{{Google Cloud Platform}(2018)}]{GoogleCloudPlatform2018}
{Google Cloud Platform} (2018) {GitHub Data}.
  \urlprefix\url{https://cloud.google.com/bigquery/public-data/github}

\bibitem[{Gousios(2013)}]{Gousios2013}
Gousios G (2013) {The GHTorrent dataset and tool suite}. In: Zimmermann T, {Di
  Penta} M, Kim S (eds) {10th International Working Conference on Mining
  Software Repositories (MSR 2013)}, IEEE, San Francisco, CA, USA, pp 233--236

\bibitem[{Hinkle et~al.(1979)Hinkle, Wiersma, and
  Jurs}]{HinkleWiersmaOthers1979}
Hinkle DE, Wiersma W, Jurs SG (1979) {Applied statistics for the behavioral
  sciences}. {Rand McNally College Publishing}, Skokie, IL, USA

\bibitem[{Juergens et~al.(2009)Juergens, Deissenboeck, Hummel, and
  Wagner}]{JuergensDeissenboeckOthers2009}
Juergens E, Deissenboeck F, Hummel B, Wagner S (2009) {Do Code Clones Matter?}
  In: Fickas S, Atlee JM, Inverardi P (eds) {31st International Conference on
  Software Engineering (ICSE 2009)}, {IEEE Computer Society}, Vancouver, BC,
  Canada, pp 485--495

\bibitem[{Lancaster and Culwin(2004)}]{LancasterCulwin2004}
Lancaster T, Culwin F (2004) {A comparison of source code plagiarism detection
  engines}. {Computer Science Education} 14(2):101--112

\bibitem[{Lehman(1980)}]{Lehman1980}
Lehman MM (1980) {Programs, life cycles, and laws of software evolution}.
  {Proceedings of the IEEE} 68(9):1060--1076

\bibitem[{Manning et~al.(2008)Manning, Raghavan, and
  Schutze}]{ManningRaghavanOthers2008}
Manning CD, Raghavan P, Schutze H (2008) {Introduction to Information
  Retrieval}. {Cambridge University Press}, New York, NY, USA

\bibitem[{Martins et~al.(2014)Martins, Fonte, Henriques, and
  Cruz}]{MartinsFonteOthers2014}
Martins VT, Fonte D, Henriques PR, Cruz Dd (2014) {Plagiarism Detection: A Tool
  Survey and Comparison}. In: Pereira MJV, Leal JP, Simoes A (eds) {3rd
  Symposium on Languages, Applications and Technologies (SLATE 2014)}, {Schloss
  Dagstuhl--Leibniz-Zentrum fuer Informatik}, Bragan{\c{c}}a, Portugal,
  {OpenAccess Series in Informatics (OASIcs)}, vol~38, pp 143--158

\bibitem[{Matthews(1975)}]{Matthews1975}
Matthews BW (1975) {Comparison of the predicted and observed secondary
  structure of T4 phage lysozyme}. {Biochimica et Biophysica Acta (BBA) --
  Protein Structure} 405(2):442--451

\bibitem[{Mens and Demeyer(2008)}]{MensDemeyer2008}
Mens T, Demeyer S (eds)  (2008) {Software Evolution}. Springer, Berlin, Germany

\bibitem[{Morrison and Murphy-Hill(2013)}]{MorrisonMurphyHill2013}
Morrison P, Murphy-Hill E (2013) {Is programming knowledge related to age? An
  exploration of Stack Overflow}. In: Zimmermann T, {Di Penta} M, Kim S (eds)
  {10th International Working Conference on Mining Software Repositories (MSR
  2013)}, IEEE, San Francisco, CA, USA, pp 69--72

\bibitem[{Nasehi et~al.(2012)Nasehi, Sillito, Maurer, and
  Burns}]{NasehiSillitoOthers2012}
Nasehi SM, Sillito J, Maurer F, Burns C (2012) {What makes a good code example?
  A study of programming Q{\&}A in StackOverflow}. In: Tonella P, {Di Penta} M,
  Maletic JI (eds) {28th IEEE International Conference on Software Maintenance
  (ICSM 2012)}, {IEEE Computer Society}, Trento, Italy, pp 25--34

\bibitem[{Parnin et~al.(2012)Parnin, Treude, Grammel, and
  Storey}]{ParninTreudeOthers2012}
Parnin C, Treude C, Grammel L, Storey MA (2012) {Crowd documentation: Exploring
  the coverage and the dynamics of API discussions on Stack Overflow}. {Georgia
  Institute of Technology, Technical Report}

\bibitem[{Ponzanelli et~al.(2013)Ponzanelli, Bacchelli, and
  Lanza}]{PonzanelliBacchelliOthers2013}
Ponzanelli L, Bacchelli A, Lanza M (2013) {Seahawk: Stack Overflow in the IDE}.
  In: Notkin D, Cheng BHC, Pohl K (eds) {35th International Conference on
  Software Engineering (ICSE 2013)}, {IEEE Computer Society}, San Francisco,
  CA, USA, pp 1295--1298

\bibitem[{Ponzanelli et~al.(2014)Ponzanelli, Mocci, Bacchelli, and
  Lanza}]{PonzanelliMocciOthers2014}
Ponzanelli L, Mocci A, Bacchelli A, Lanza M (2014) {Understanding and
  classifying the quality of technical forum questions}. In: Wong WE, McMillin
  B (eds) {14th International Conference on Quality Software (QSIC 2014)},
  IEEE, Allen, TX, USA, pp 343--352

\bibitem[{Powers(2011)}]{Powers2011}
Powers DM (2011) {Evaluation: From precision, recall and F-measure to ROC,
  informedness, markedness and correlation}. {Journal of Machine Learning
  Technologies} 2(1):37--63

\bibitem[{Roy et~al.(2009)Roy, Cordy, and Koschke}]{RoyCordyOthers2009}
Roy CK, Cordy JR, Koschke R (2009) {Comparison and evaluation of code clone
  detection techniques and tools: A qualitative approach}. {Science of Computer
  Programming} 74(7):470--495

\bibitem[{Schleimer et~al.(2003)Schleimer, Wilkerson, and
  Aiken}]{SchleimerWilkersonOthers2003}
Schleimer S, Wilkerson DS, Aiken A (2003) {Winnowing: Local algorithms for
  document fingerprinting}. In: Halevy AY, Ives ZG, Doan A (eds) {2003 ACM
  SIGMOD International Conference on Management of Data (SIGMOD 2003)}, ACM,
  San Diego, CA, USA, pp 76--85

\bibitem[{Silvestri et~al.(2015)Silvestri, Yang, Bozzon, and
  Tagarelli}]{SilvestriYangOthers2015}
Silvestri G, Yang J, Bozzon A, Tagarelli A (2015) {Linking Accounts across
  Social Networks: The Case of StackOverflow, GitHub and Twitter}. In: Armano
  G, Bozzon A, Giuliani A (eds) {1st International Workshop on Knowledge
  Discovery on the WEB (KDWeb 2015)}, CEUR-WS.org, Cagliari, Italy, {CEUR
  Workshop Proceedings}, pp 41--52

\bibitem[{Spearman(1904)}]{Spearman1904}
Spearman C (1904) {The proof and measurement of association between two
  things}. {American Journal of Psychology} 15(1):72--101

\bibitem[{{Stack Exchange Community
  Wiki}(2018-02-27)}]{StackExchangeCommunityWiki20180227}
{Stack Exchange Community Wiki} (2018-02-27) {Database schema documentation for
  the public data dump and SEDE}.
  \urlprefix\url{https://meta.stackexchange.com/a/2678}

\bibitem[{{Stack Exchange Inc}(2017)}]{StackExchangeInc2017b}
{Stack Exchange Inc} (2017) {Stack Exchange Data Dump 2017-12-01}.
  \urlprefix\url{https://archive.org/details/stackexchange/}

\bibitem[{{Stack Exchange Inc}(2018)}]{StackExchangeInc2018}
{Stack Exchange Inc} (2018) {Markdown help}.
  \urlprefix\url{https://stackoverflow.com/editing-help}

\bibitem[{{Stack Overflow Meta}(2018)}]{StackOverflowMeta2018}
{Stack Overflow Meta} (2018) {How to handle code clones on Stack Overflow?}
  \urlprefix\url{https://meta.stackoverflow.com/q/375761}

\bibitem[{Subramanian and Holmes(2013)}]{SubramanianHolmes2013}
Subramanian S, Holmes R (2013) {Making sense of online code snippets}. In:
  Zimmermann T, {Di Penta} M, Kim S (eds) {10th International Working
  Conference on Mining Software Repositories (MSR 2013)}, IEEE, San Francisco,
  CA, USA, pp 85--88

\bibitem[{Thummalapenta et~al.(2010)Thummalapenta, Cerulo, Aversano, and {Di
  Penta}}]{ThummalapentaCeruloOthers2010}
Thummalapenta S, Cerulo L, Aversano L, {Di Penta} M (2010) {An empirical study
  on the maintenance of source code clones}. {Empirical Software Engineering}
  15(1):1--34

\bibitem[{Treude and Robillard(2016)}]{TreudeRobillard2016}
Treude C, Robillard MP (2016) {Augmenting API Documentation with Insights from
  Stack Overflow}. In: Dillon L, Visser W, Williams L (eds) {38th International
  Conference on Software Engineering (ICSE 2016)}, ACM, Austin, TX, USA, pp
  392--403

\bibitem[{Treude and Robillard(2017)}]{TreudeRobillard2017}
Treude C, Robillard MP (2017) {Understanding Stack Overflow Code Fragments}.
  In: Mei H, Zhang L, Zimmermann T (eds) {2017 IEEE International Conference on
  Software Maintenance and Evolution (ICSME 2017)}, {IEEE Computer Society},
  Shanghai, China, pp 509--513

\bibitem[{Treude et~al.(2011)Treude, Barzilay, and
  Storey}]{TreudeBarzilayOthers2011}
Treude C, Barzilay O, Storey MAD (2011) {How do programmers ask and answer
  questions on the web?} In: Taylor RN, Gall HC, Medvidovic N (eds) {33rd
  International Conference on Software Engineering (ICSE 2011)}, ACM, Waikiki,
  Honolulu, pp 804--807

\bibitem[{Vasilescu et~al.(2012)Vasilescu, Capiluppi, and
  Serebrenik}]{VasilescuCapiluppiOthers2012}
Vasilescu B, Capiluppi A, Serebrenik A (2012) {Gender, Representation and
  Online Participation: A Quantitative Study of StackOverflow}. In: Aberer K,
  Flache A, Jager W, Liu L, Tang J, Gueret C (eds) {4th International
  Conference on Social Informatics (SocInfo 2012)}, Springer, Lausanne,
  Switzerland, {Lecture Notes in Computer Science}, pp 332--338

\bibitem[{Vasilescu et~al.(2013)Vasilescu, Filkov, and
  Serebrenik}]{VasilescuFilkovOthers2013}
Vasilescu B, Filkov V, Serebrenik A (2013) {StackOverflow and GitHub:
  Associations between Software Development and Crowdsourced Knowledge}. In:
  Chang LW, Srivastava J, Zhan J (eds) {2013 International Conference on Social
  Computing (SocialCom 2013)}, {IEEE Computer Society}, Washington, DC, USA, pp
  188--195

\bibitem[{Wang et~al.(2013)Wang, {Lo David}, and Jiang}]{WangLoDavidOthers2013}
Wang S, {Lo David}, Jiang L (2013) {An empirical study on developer
  interactions in StackOverflow}. In: Shin SY, Maldonado JC (eds) {28th Annual
  ACM Symposium on Applied Computing (SAC 2013)}, ACM, Coimbra, Portugal, pp
  1019--1024

\bibitem[{Wilcoxon(1945)}]{Wilcoxon1945}
Wilcoxon F (1945) {Individual comparisons by ranking methods}. {Biometrics}
  1(6):80--83

\bibitem[{Xia et~al.(2017)Xia, Bao, Lo, Kochhar, Hassan, and
  Xing}]{XiaBaoOthers2017}
Xia X, Bao L, Lo D, Kochhar PS, Hassan AE, Xing Z (2017) {What do developers
  search for on the web?} {Empirical Software Engineering} 22(6):3149--3185

\bibitem[{Yang et~al.(2016)Yang, Hussain, and Lopes}]{YangHussainOthers2016}
Yang D, Hussain A, Lopes CV (2016) {From Query to Usable Code: An Analysis of
  Stack Overflow Code Snippets}. In: Kim M, Robbes R, Bird C (eds) {13th
  International Conference on Mining Software Repositories (MSR 2016)}, ACM,
  Austin, TX, USA, pp 391--402

\bibitem[{Yang et~al.(2017)Yang, Martins, Saini, and
  Lopes}]{YangMartinsOthers2017}
Yang D, Martins P, Saini V, Lopes CV (2017) {Stack Overflow in Github: Any
  Snippets There?} In: Gonzalez-Barahona JM, Hindle A, Tan L (eds) {14th
  International Conference on Mining Software Repositories (MSR 2017)}, {IEEE
  Computer Society}, Buenos Aires, Argentina, pp 280--290

\bibitem[{Yang et~al.(2014)Yang, Hauff, Bozzon, and
  Houben}]{YangHauffOthers2014}
Yang J, Hauff C, Bozzon A, Houben GJ (2014) {Asking the right question in
  collaborative Q{\&}A systems}. In: Ferres L, Rossi G, Almeida VAF, Herder E
  (eds) {25th ACM Conference on Hypertext and Social Media (HT 2014)}, ACM,
  Santiago, Chile, pp 179--189

\end{thebibliography}

\end{document}